\documentclass[review]{elsarticle}
\usepackage{framed}
\usepackage{multicol}
\usepackage{subcaption}
\usepackage{graphicx}
\usepackage{amsmath}
\usepackage{color,soul}
\usepackage{soul}
\usepackage{xcolor}
\usepackage{rotating}
\usepackage{comment}
\usepackage{multirow}

\usepackage{nomencl}
\makenomenclature

\usepackage{lineno,hyperref}

\usepackage{geometry}
\makeatletter
\def\ps@pprintTitle{%
	\let\@oddhead\@empty
	\let\@evenhead\@empty
	\def\@oddfoot{}%
	\let\@evenfoot\@oddfoot}
\makeatother

\begin{document}

\begin{frontmatter}

\title{Fourier series based modeling of the dynamics of inclined closed loop buoyancy driven heat exchangers with conjugate effect}

\author{Akhil Dass, Sateesh Gedupudi \footnote{Corresponding author. Tel.: +91 44 2257 4721, Email: sateeshg@iitm.ac.in}}
\address{Heat Transfer and Thermal Power Laboratory, Department of Mechanical Engineering, IIT Madras, Chennai 600036, India}

\begin{abstract}
The modelling of the dynamics of inclined closed loop buoyancy driven heat exchangers with inclusion of the wall conduction effect at the heat exchanger is presented in the current study. A Coupled Natural Circulation Loop (CNCL) is an ideal system for studying the closed loop buoyancy driven heat exchanger. The modelling utilises a Fourier series based approach to develop a 1-D model which is then verified with the 3-D CFD studies of the respective cases. A good agreement is observed with the 3-D CFD data, which demonstrates the suitability of the 1-D model for transient behaviour prediction. The non-dimensional numbers and thermal coupling sensitivity coefficients which govern the dynamics of the CNCL are identified and an appropriate parametric study is conducted. Results show that the wall conduction and inclination have a significant effect on the transient behaviour of the CNCL system. A jump in the heat transfer coefficient with variation in the inclination of the Conjugate CNCL system is observed. The 1-D model is also able to capture the flow direction reversal with change in the inclination of the Conjugate CNCL system for zero flow field initial conditions.
\end{abstract}

\begin{keyword}
Fourier series, natural circulation loop, heat exchanger, conjugate effect,heat transfer jump.
\end{keyword}

\end{frontmatter}

\nolinenumbers

\section{Introduction}

Natural circulation is a buoyancy-driven flow within a closed fluid-filled conduit which is initiated by thermal excitation in the presence of gravitational field (body force field in general). The system which employs this phenomenon is called as a Natural Circulation Loop (NCL) and is extensively used in many applications considering that it does not need any external power to drive the fluid through the loop and due to its lack of moving components. It can also be used as a passive safety device. It finds applications in many engineering systems such as solar heaters, electronic chip cooling, nuclear power generation, geothermal energy extraction etc. \cite{basu2014}.
The present work focuses primarily on Coupled Natural Circulation Loop (CNCL) systems, which comprises of two NCLs thermally linked to each other at the common heat exchange location. A CNCL is an ideal system to understand the dynamics of a closed loop buoyancy driven heat exchanger (both hot fluid and cold fluid sides are propelled by buoyancy forces). A detailed study of such systems with a 1-D modelling approach for the most basic geometry has been performed by Dass and Gedupudi \cite{dass2019} for pure convective flow systems. The real systems encountered in practice always have multiple heat transfer mechanisms influencing the dynamics of the system. Thus, it is necessary to account for the other heat transfer mechanism namely conduction (in the pipe walls) apart from the convective flow within the loop which play a vital role in the dynamics of the circulation loop system. 
To model the conjugate heat transfer in the CNCL system, a Fourier based 1-D modelling approach is utilised in the current work. The 1-D Fourier series based modelling of an NCL was developed by Hart \cite{hart1984} for a toroidal NCL system. Davis and Roppo \cite{davis1987} employed the same methodology to study a CNCL with toroid component loops and a point contact. Rodriguez and Van Vleck \cite{rodriguez1998} extended the method further by making it suitable for a generic NCL geometry and consideirng the fluid axial conduction. Fichera and Pagano \cite{fichera2003} utilised the generic model proposed and employed it to model the rectangular NCL system. To further extend its practical relevance, Dass and Gedupudi \cite{dass2019} employed the Fourier series based modelling approach to model CNCL with rectangular loops having a non-point contact at the section where the constituent loops are thermally coupled. 
From the literature review, it can be observed that the 1-D modelling approach based on the Fourier series has been significantly extended but without considering the wall conduction effects. The influence of the conjugate heat transfer on the circulation systems has been emphasized by Misale \cite{misale2014} as an open question. Benne and Homan \cite{benne2008},\cite{benne2009} conducted studies on a single NCL system coupled to an external thermal storage. The present study models a CNCL system with rectangular component NCLs thermally linked to each other at the common heat exchange section with the wall thickness effect included. This system is referred to as a conjugate CNCL system from here on and is utilised to demonstrate the conjugate effects on the system dynamics. Misale et al. \cite{misale2000} conducted a 2-D CFD study to investigate the influence of conjugate effects on NCL systems. A finite difference approach was utilised to obtain the transient system behaviour and a good match was reported between the model prediction and the experimental data. Another work conducted by Misale et al. \cite{misale2005} reports the effect of inclination and a systematic parametric study on the influencing parameters in a conjugate NCL system. The work also reports the influence of wall conduction on the stability of the NCL system behavior. Thus, from the available literature it is clearly evident that the inclusion of conjugate effects makes the developed model more realistic, and so the present study focuses on modeling the conjugate effects in CNCL systems.

The effect of inclination on the Conjugate CNCL system is also an important aspect of the present study. The inclination considered in the current study is confined to the plane of the Conjugate CNCL. Ramos et al. \cite{ramos1990} conducted a 2-D steady state CFD study to investigate the influence of inclination on a square NCL system. The inclination of the NCL had an influence on the streamline flow patterns and the existence of multiple steady state solutions with respect to the inclination was also demonstrated. Basu at al. \cite{basu2013influence} and Krishnani and Basu \cite{krishnani2017} studied the effect of inclination on the stability of the NCL system. Introducing a small tilt angle was reported to have a significant stabilizing effect even at large power input condition. A detailed 2-D CFD study of inclined NCL and CNCL systems conducted by Dass and Gedupudi \cite{dass2020} reports a  jump in the heat transfer coefficient with change in inclination. Another motive of the present study is to evaluate the occurrence of similar behaviour in conjugate CNCL systems. 

CFD investigation of natural circulation systems has become a standard practice as observed from the recent thorough investigation of simple NCL systems carried out by Hashemi-Tilehnoee et al. \cite{hashemi2019simulation} and Dai et al. \cite{dai2020heat}, and from the study of complex systems which employ natural circulation such as passive decay heat removal systems conducted by Minocha et al. \cite{minocha2015numerical}. The present study also employs a 3-D CFD study to verify the developed 1-D  model of the inclined conjugate CNCL systems and to investigate the physics of the system.

The objectives of the present study can be summarised as follows:
\begin{enumerate}
    \item To develop a 1-D model of CNCL system incorporating the conjugate and inclination effects.
    \item To identify and characterize the newly identified non-dimensional numbers which govern the Conjugate CNCL system behaviour.
    \item To conduct a 3-D CFD study of Conjugate CNCL system for verification of the 1-D model.
    \item To determine the effect of inclination on the Conjugate CNCL system behaviour and to verify the occurrence of jump in the heat transfer coefficient with inclination.
    \item To conduct a thorough parametric study of the inclined conjugate CNCL system.
\end{enumerate}

\section{Modeling of inclined Conjugate CNCL systems}

The current sections presents the 1-D modeling methodology employed to capture the conjugate effects in inclined CNCL systems. The following assumptions are employed to model the system:

\begin{enumerate}
    \item The Boussinesq approximation is utilised to model the buoyancy forces.
    \item One-dimensional governing equations are employed.
    \item The axial conduction effects along the flow direction are accounted for both the solid and fluid domains.
    \item The wall is assumed to be lumped in the direction normal to the fluid flow.
    \item Viscous dissipation term in the energy equation is neglected.
\end{enumerate}

\subsection{Governing equations of the inclined Conjugate CNCL system}

\begin{figure}[!htb]
    \centering
    \includegraphics[width=\linewidth]{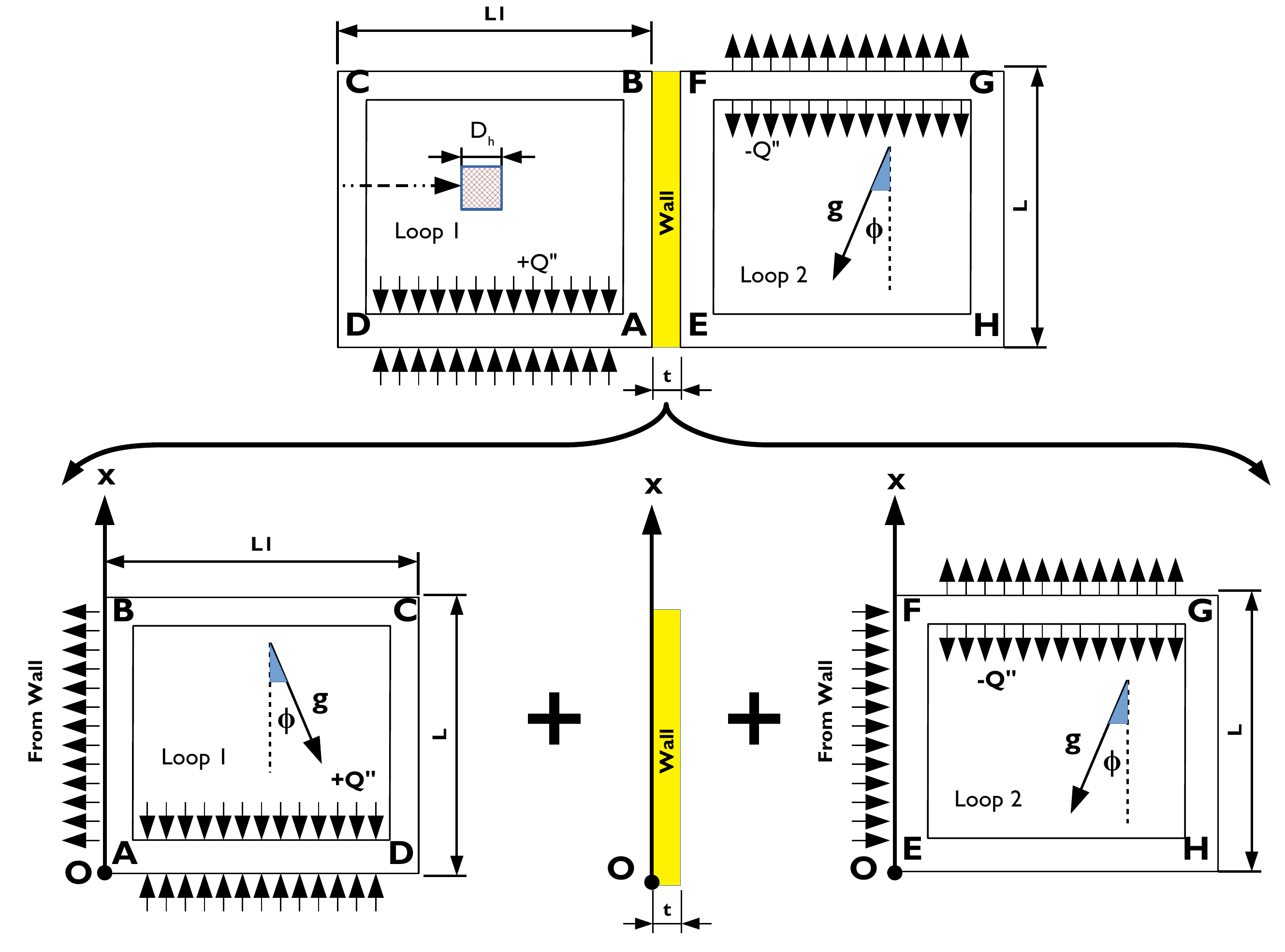}
    \caption{Schematic of the modeling approach. ABCD and EFGH indicate the direction which is considered positive in each loop and O is the origin}
    \label{Modeling Conjugate CNCL}
\end{figure}

The governing equations can be derived from the force and energy balance on a representative element of Loop 1 or Loop 2 of the conjugate CNCL system. Detailed derivations of momentum and energy equations for loop 1 (equations (1), (2)) and those for loop 2 (equations (4), (5)) are available in \cite{dass2019}. Equation (3) represents the wall conduction at the heat exchanger section. The coordinate system used for the derivation of the governing equations and the schematic of the modeling approach utilised is shown in Figure\ref{Modeling Conjugate CNCL}.

\begin{equation}
    \rho_1 \frac{d \omega_1}{dt} + \frac{4 \tau_1}{D_h} =
    \frac{\rho_1 g \beta_1}{2(L+L1)} \oint (T_1-T_0) f(x) dx - \frac{NK\rho_1 \omega_1 ^2}{4(L+L1)}
\end{equation}

\begin{equation}
    \frac{\partial T_1}{\partial t} + \omega_1\frac{\partial T_1}{\partial x} = \frac{4Q^{\prime \prime}}{\rho_1 C_{p,1} D_h} - \frac{U \lambda (x)}{\rho_1 C_{p,1} D_h} (T_1 - T_w) + \alpha_1\frac{\partial^2 T_1}{\partial x^2}
\end{equation}

\begin{equation}
    \frac{\partial T_w}{\partial t}= \frac{U \lambda (x)}{\rho_w C_{p,w} t_w} (T_1 +T_2 - 2T_w) + \lambda (x)\alpha_w\frac{\partial^2 T_w}{\partial x^2}
\end{equation}

\begin{equation}
    \rho_2 \frac{d \omega_2}{dt} + \frac{4 \tau_2}{D_h} =
    \frac{\rho_2 g \beta_2}{2(L+L1)} \oint (T_2-T_0) g(x) dx - \frac{NK\rho_1 \omega_2 ^2}{4(L+L1)}
\end{equation}

\begin{equation}
    \frac{\partial T_2}{\partial t} + \omega_2\frac{\partial T_2}{\partial x} = \frac{4Q^{\prime \prime}}{\rho_2 C_{p,2} D_h} - \frac{U \lambda (x)}{\rho_2 C_{p,2} D_h} (T_w - T_2) + \alpha_2\frac{\partial^2 T_2}{\partial x^2}
\end{equation}

where,

\begin{equation}
    \tau_1=\frac{\rho_1 \omega_1^2}{2}\bigg( \frac{b}{Re_1} \bigg)^d
\end{equation}

\begin{equation}
    \tau_2=\frac{\rho_2 \omega_2^2}{2}\bigg( \frac{b}{Re_2} \bigg)^d
\end{equation}

\begin{equation}
    K=\frac{800}{Re_i} +0.14 \bigg(1 + \frac{4}{(D_h)^{0.3}} \bigg)
\end{equation}

\begin{equation}
    f(x)=\left\{
        \begin{array}{llll}
            cos(\phi), & \quad 0< x < 0 \\
            -sin(\phi), & \quad L<x < L+L1 \\
            -cos(\phi), & \quad L+L1<x< 2L+L1 \\
            sin(\phi), & \quad 2L+L1 < x< 2(L+L1)
        \end{array}
    \right.
\end{equation}

\begin{equation}
    g(x)=\left\{
        \begin{array}{llll}
            cos(\phi), & \quad 0< x < 0 \\
            sin(\phi), & \quad L<x < L+L1 \\
            -cos(\phi), & \quad L+L1<x< 2L+L1 \\
            -sin(\phi), & \quad 2L+L1 < x< 2(L+L1)
        \end{array}
    \right.
\end{equation}

\begin{equation}
    \lambda(x)=\left\{
        \begin{array}{llll}
            1, & \quad 0< x < 0 \\
            0, & \quad L<x < L+L1 \\
            0, & \quad L+L1<x< 2L+L1 \\
            0, & \quad 2L+L1 < x< 2(L+L1)
        \end{array}
    \right.
\end{equation}

\begin{equation}
    h_1(x)=\left\{
        \begin{array}{llll}
            0, & \quad 0< x < 0 \\
            0, & \quad L<x < L+L1 \\
            0, & \quad L+L1<x< 2L+L1 \\
            1, & \quad 2L+L1 < x< 2(L+L1)
        \end{array}
    \right.
\end{equation}

\begin{equation}
    h_2(x)=\left\{
        \begin{array}{llll}
            0, & \quad 0< x < 0 \\
            1, & \quad L<x < L+L1 \\
            0, & \quad L+L1<x< 2L+L1 \\
            0, & \quad 2L+L1 < x< 2(L+L1)
        \end{array}
    \right.
\end{equation}

$f(x)$ and $g(x)$ are functions which represent the inclinations w.r.t. gravity of Loop 1 and Loop 2, respectively. $\lambda(x)$ is a function which represents the location of interaction between the fluid of Loop 1, the heat exchanger wall and the fluid of Loop 2. $h_1(x)$ and $h_2(x)$ are functions which represent the locations of the heating and cooling sections on Loop 1 and Loop 2 respectively. Derivation of Equation (3) can be found in the appendix section.

\subsection{Initial conditions of the tilted Conjugate CNCL system}

\begin{equation}
    \omega_1 (t=0)=0; \; \omega_2 (t=0)=0 
\end{equation}

\begin{equation}
    T_1 (x,t=0)=T_0; \; T_2 (x,t=0)=T_0; \: T_w (x,t=0)=T_0
\end{equation}

\subsection{Boundary conditions at the heat exchanger wall}

\begin{equation}
    \frac{\partial T_w}{\partial x}\bigg|_{x=0} =0; \; \frac{\partial T_w}{\partial x}\bigg|_{x=L} =0;
\end{equation}

\subsection{Non-dimensional governing equations of the conjugate CNCL system}

Let us consider the non-dimensional variables as follows:

\begin{equation*}
    \theta_1=\frac{T_1-T_0}{\Delta T_1}, \; \theta_2=\frac{T_2-T_0}{\Delta T_2}, \;
    \theta_w=\frac{T_w-T_0}{\Delta T_w}, \;
    \zeta=\frac{t}{t_0}, \; s=\frac{x}{x_0}, \; Re_1=\frac{\omega_1 D_h}{\nu_1}, \; Re_2=\frac{\omega_2 D_h}{\nu_2}
\end{equation*}

where,

\begin{equation*}
\Delta T_1=\frac{4 Q^{\prime \prime}t_0}{\rho_1 C_{p,1} D_h}, \; \Delta T_2=\frac{4 Q^{\prime \prime}t_0}{\rho_2 C_{p,2} D_h}, \; \Delta T_w=\frac{4 Q^{\prime \prime}t_0}{\rho_w C_{p,w} t_w}, \;
t_0=\frac{x_0 D_h}{\nu_1}, \; x_0=(L+L1)
\end{equation*}

Substituting the fore listed non-dimensional variables into equations (1-5) and considering

\begin{equation*}
    Gr_i=\frac{g \beta_i \Delta T_i x_0 D_h t_0}{2(L+L1) \nu}, \;
    Fo_i=\frac{\alpha_i t_0}{x_0^2}, \; Fo_w=\frac{\alpha_w t_0}{x_0^2}, \; St=\frac{Ut_0}{\rho_i C_{p,i} D_h}  
\end{equation*}

\begin{equation*}
    Co_1=\frac{2b\nu_1 t_0}{D_h^2}, \; Co_2=\frac{\Delta T_w}{\Delta T_1},\; Co_3=\frac{\Delta T_2}{\Delta T_1},\; Co_4=\frac{\Delta T_w}{\Delta T_2}
\end{equation*}

we obtain the following non-dimensional equations which govern the dynamics of the tilted Conjugate CNCL system which are represented as follows:

\begin{equation}
\frac{d Re_1}{d\zeta} + Co_1\; Re_1^{(2-d)} = Gr_1\oint \theta_1 \;f(s)ds - \frac{NK}{4}Re_1^2    
\end{equation}

\begin{equation}
    \frac{\partial \theta_1}{\partial \zeta} +Re_1\frac{\partial \theta_1}{\partial s} = h_1(s) -\lambda(s) St_1 (\theta_1 - Co_2\; \theta_w) + Fo_1\frac{\partial^2 \theta_1}{\partial s^2}
\end{equation}

\begin{equation}
    \frac{\partial \theta_w}{\partial \zeta} =\lambda(s) St_1 (\theta_1 + Co_3\;\theta_2 -2 \;Co_2\; \theta_w) + \lambda_s(s)Fo_w\frac{\partial^2 \theta_w}{\partial s^2}
\end{equation}

\begin{equation}
\frac{d Re_2}{d\zeta} + \frac{\nu_2}{\nu_1}\;Co_1\;Re_2^{(2-d)} = Gr_2\oint \theta_2 \;g(s)ds - \frac{\nu_2}{\nu_1}\;\frac{NK}{4}Re_2^2    
\end{equation}

\begin{equation}
    \frac{\partial \theta_2}{\partial \zeta} +\frac{\nu_2}{\nu_1}\;Re_2\frac{\partial \theta_2}{\partial s} = h_2(s) +\lambda(s) St_2 (Co_4\;\theta_w - \theta_2) + Fo_2\frac{\partial^2 \theta_2}{\partial s^2}
\end{equation}

\subsection{Physical significance of $Co_1$, $Co_2$, $Co_3$ and $Co_4$  }

The non-dimensional number $Co_1$ is flow resistance coefficient as defined by Dass and Gedupudi (2019) \cite{dass2019}. It denotes the resistance encountered by the fluid. The non-dimensinal number $Co_3$ is the thermal coupling sensitivity coefficient as defined by  Dass and Gedupudi (2019) \cite{dass2019}. It denotes the strength and sensitivity of coupling between the fluids of Loop 1 and Loop 2.

$Co_2$ is defined as follows:

\begin{equation}
    Co_2=\frac{\Delta T_w}{\Delta T_1}=\frac{\rho_1 C_{p,1} D_h}{\rho_w C_{p,w} t_w}
\end{equation}

From equation 18 we observe that the strength of the coupling between the fluid in Loop 1 and the common heat exchanger wall will be  affected by the magnitude of $Co_2$. Thus, it maybe inferred that $Co_2$ is the thermal sensitivity coupling coefficient between fluid of Loop 1 and the heat exchanger wall.

$Co_4$ is defined as follows:

\begin{equation}
    Co_4=\frac{\Delta T_w}{\Delta T_2}=\frac{\rho_2 C_{p,2} D_h}{\rho_w C_{p,w} t_w}
\end{equation}

From equation 21 we observe that the strength of the coupling between the fluid in Loop 2 and the common heat exchanger wall will be  affected by the magnitude of $Co_4$. Thus, it maybe inferred that $Co_4$ is the thermal sensitivity coupling coefficient between fluid of Loop 2 and the heat exchanger wall.

From the above analysis, we can infer that incorporating the conjugate effects in the CNCL system leads to more complex interaction between the fluids of Loop 1, Loop 2 and the common heat exchange wall resulting in three thermal coupling sensitivity  coefficients, namely $Co_2$, $Co_3$ and $Co_4$.

\section{Solution methodology}

\subsection{Simplification of the partial differential equations}

To solve the non-dimensional Partial Differential Equation (PDE) of the conjugate CNCL system we employ a Fourier series-based approach to convert the PDE’s to a set of ordinary differential equations. The following functions are represented using the Fourier series:

\begin{equation*}
    f(s)=\sum_{k=-\infty}^{\infty} f_k e^{(i\pi k s)}, \; g(s)=\sum_{k=-\infty}^{\infty} g_k e^{(i\pi k s)}, \;
\lambda(s)=\sum_{k=-\infty}^{\infty} \lambda_k e^{(i\pi k s)}, \;
\end{equation*}

\begin{equation*}
    h_1(s)=\sum_{k=-\infty}^{\infty} h_{1_k} e^{(i\pi k s)}, \;
    h_2(s)=\sum_{k=-\infty}^{\infty} h_{2_k} e^{(i\pi k s)}
\end{equation*}

\begin{equation*}
    \theta_1(s,\zeta)=\sum_{k=-\infty}^{\infty} \theta_{1_k}(\zeta) e^{(i\pi k s)}, \;
    \theta_2(s,\zeta)=\sum_{k=-\infty}^{\infty} \theta_{2_k}(\zeta) e^{(i\pi k s)}, \;
    \theta_w(s,\zeta)=\sum_{k=-\infty}^{\infty} \theta_{w_k}(\zeta) e^{(i\pi k s)}
\end{equation*}

Substituting the above-mentioned Fourier series into equations (17) to (21) provides us the general stencil which represents the conjugate CNCL as a system of ODEs. The stencil is represented as follows:

\begin{equation}
\frac{d Re_1}{d\zeta} + Co_1\; Re_1^{(2-d)} = Gr_1\sum_{k=-\infty}^{\infty} \theta_{1_k} f_{(-k)} - \frac{NK}{4}Re_1^2    
\end{equation}

\begin{equation}
    \frac{d \theta_{1_k}}{d \zeta} + ik\pi Re_1 \; \theta_{1_k} = h_{1_k} - St_1 \sum_{m=-\infty}^{\infty} \lambda_{(k-m)}(\theta_{1_m} - Co_2\; \theta_{w_m}) - (k \pi)^2 Fo_1 \; \theta_{1_k}
\end{equation}

\begin{equation}
    \frac{d \theta_{w_k}}{d \zeta}  = 4 St_1 \sum_{m=-\infty}^{\infty} \lambda_{(k-m)} \bigg(  \theta_{1_m} - \bigg( 2Co_2  + (m \pi)^2 \frac{Fo_w}{St_1} \bigg) \theta_{w_m} + Co_3 \; \theta_{2_m}  \bigg) 
\end{equation}

\begin{equation}
\frac{d Re_2}{d\zeta} + \frac{\nu_2}{\nu_1}\;Co_1\;Re_2^{(2-d)} = Gr_2\sum_{k=-\infty}^{\infty} \theta_{2_k} g_{(-k)} - \frac{\nu_2}{\nu_1}\;\frac{NK}{4}Re_2^2    
\end{equation}

\begin{equation}
    \frac{d \theta_{2_k}}{d \zeta} + ik\pi\frac{\nu_2}{\nu_1} Re_2 \; \theta_{2_k} = h_{2_k} + St_2 \sum_{m=-\infty}^{\infty} \lambda_{(k-m)}(Co_4\;\theta_{w_m} - \theta_{2_m}) - (k \pi)^2 Fo_2 \; \theta_{2_k}
\end{equation}

The above mentioned set of stencils from equations (24) to (28) represent the complete conjugate CNCL system where $k \in (-\infty,\infty)$. For the current study the number of Fourier nodes has been restricted to three which is found to be adequate to represent the dynamics of the system. This is further justified by the node independence test carried out in the upcoming section.

The modelling of the Conjugate CNCL system is simplified if we utilise same fluid in both the loops. This leads to the following simplifications:

\begin{equation*}
    Gr_1=Gr_2=Gr, \; St_1=St_2=St, \; Fo_1=Fo_2=Fo, \; Co_1=Co_A 
\end{equation*}

\begin{equation*}
    \frac{\nu_2}{\nu_1}=1, \; \Delta T_1 =\Delta T_2 \implies Co_2=Co_4=Co_B, \; Co_3=1
\end{equation*}

Incorporating the above mentioned simplifications and expanding the CNCL system stencil with three Fourier nodes ($k \in (-3,3)$) and separating them into real and imaginary parts results in the following set of ODEs, represented by equations (29) to (36) as follows:

\begin{equation}
\frac{d Re_1}{d\zeta} + Co_A\; Re_1^{(2-d)} = Gr\sum_{k=-3}^{3} \theta_{1_k} f_{(-k)} - \frac{NK}{4}Re_1^2    
\end{equation}

\begin{equation}
    \frac{d [\theta_{1_k}]^R}{d \zeta} - ik\pi Re_1 \; [\theta_{1_k}]^I = [h_{1_k}]^R - \bigg[St_1 \sum_{m=-3}^{3} \lambda_{(k-m)}(\theta_{1_m} - Co_B\; \theta_{w_m})\bigg]^R - (k \pi)^2 Fo \; [\theta_{1_k}]^R
\end{equation}

\begin{equation}
    \frac{d [\theta_{1_k}]^I}{d \zeta} + ik\pi Re_1 \; [\theta_{1_k}]^R = [h_{1_k}]^I - \bigg[St_1 \sum_{m=-3}^{3} \lambda_{(k-m)}(\theta_{1_m} - Co_B\; \theta_{w_m})\bigg]^I - (k \pi)^2 Fo \; [\theta_{1_k}]^I
\end{equation}

\begin{equation}
    \frac{d [\theta_{w_k}]^R}{d \zeta}  = 4 St \bigg[ \sum_{m=-3}^{3} \lambda_{(k-m)} \bigg(  \theta_{1_m} - \bigg( 2Co_2  + (m \pi)^2 \frac{Fo_w}{St} \bigg) \theta_{w_m} + Co_3 \; \theta_{2_m}  \bigg) \bigg]^R
\end{equation}

\begin{equation}
    \frac{d [\theta_{w_k}]^I}{d \zeta}  = 4 St \bigg[ \sum_{m=-3}^{3} \lambda_{(k-m)} \bigg(  \theta_{1_m} - \bigg( 2Co_2  + (m \pi)^2 \frac{Fo_w}{St} \bigg) \theta_{w_m} + Co_3 \; \theta_{2_m}  \bigg) \bigg]^I
\end{equation}

\begin{equation}
\frac{d Re_2}{d\zeta} + Co_A\;Re_2^{(2-d)} = Gr\sum_{k=-3}^{3} \theta_{2_k} g_{(-k)} - \frac{NK}{4}Re_2^2    
\end{equation}

\begin{equation}
    \frac{d [\theta_{2_k}]^R}{d \zeta} - ik\pi Re_2 \; [\theta_{2_k}]^I = [h_{2_k}]^R - \bigg[St \sum_{m=-3}^{3} \lambda_{(k-m)}(Co_B\;\theta_{w_m} - \theta_{2_m})\bigg]^R - (k \pi)^2 Fo \; [\theta_{2_k}]^R
\end{equation}

\begin{equation}
    \frac{d [\theta_{2_k}]^I}{d \zeta} + ik\pi Re_2 \; [\theta_{2_k}]^R = [h_{2_k}]^I - \bigg[St \sum_{m=-3}^{3} \lambda_{(k-m)}(Co_B\;\theta_{w_m} - \theta_{2_m})\bigg]^I - (k \pi)^2 Fo \; [\theta_{2_k}]^I
\end{equation}

The superscript ‘R’ is used to denote the real part of the Fourier node and superscript ‘I’ is used to denote the imaginary part of the Fourier node. Equations $29-36$ represent the stencil of the ODEs to be used for determining the transient behaviour of the Conjugate CNCL system.

\subsection{Initial conditions of the non-dimensional ODE system}

The non-dimensional initial conditions of the Conjugate CNCL system are:

\begin{equation*}
    Re_1(\zeta=0)=0, \; Re_2(\zeta=0)=0
\end{equation*}

\begin{equation*}
    \theta_1(s,\zeta=0)=0, \; \theta_2(s,\zeta=0)=0, \; \theta_w(s,\zeta=0)=0
\end{equation*}

The initial conditions of non-dimensional temperature ($\theta_1$) are used to determine the initial conditions of the Fourier coefficient of $\theta_1$ ($\theta_{1_k}$). The coefficient of the Fourier series $\theta_1(s,\zeta)=\sum_{k=-\infty}^{\infty} \theta_{1_k}(\zeta) e^{ik \pi s}$ is given by:

\begin{equation}
    \theta_{1_k}(\zeta)=\frac{1}{2} \oint \theta_1(s,\zeta) e^{-ik \pi s} ds
\end{equation}

Since $\theta_1(s,\zeta=0)=0$ for all $\zeta=0$ from equation (37) we obtain that $\theta_{1_k}(\zeta=0)=0$ for all $k$. Similarly, the initial conditions of $\theta_{2_k}$ and $\theta_{w_k}$ are determined and are found to be $\theta_{2_k}(\zeta=0)=0$, $\theta_{w_k}(\zeta=0)=0$ for all $k$.

\subsection{Incorporating the boundary condition on the heat exchanger wall for the non-dimensional ODE system}

The non-dimensional boundary conditions on the heat exchanger wall are:

\begin{equation}
    \frac{\partial \theta_w (s,\zeta)}{\partial s}\bigg|_{s=0} =0 
\end{equation}

\begin{equation}
    \frac{\partial \theta_w (s,\zeta)}{\partial s}\bigg|_{s=\frac{L}{x_0}} =0 
\end{equation}

For the present study the aspect ratio ($L/L1$) is set to unity to simplify the calculations. This implies that $L=L1$ and $\frac{L}{x_0}=0.5$. We use the truncated Fourier series to represent the temperature distribution in the wall given by:

\begin{equation}
    \theta_w(s,\zeta)=\sum_{k=-3}^{3} \theta_{w_k}(\zeta) e^{ik \pi s}
\end{equation}

and

\begin{equation}
    \theta_{w_k}(\zeta)= \theta_{w_k}^R(\zeta) + i \theta_{w_k}^I(\zeta)
\end{equation}

Substituting equations (40) and (41) into equation (38) yields:

\begin{equation}
    \frac{\partial }{\partial s} \bigg( \sum_{k=-3}^{3} (\theta_{w_k}^R(\zeta) + i \theta_{w_k}^I(\zeta)) \bigg) \bigg|_{s=0} =0 
\end{equation}

Simplifying equation (42) and separating the real and imaginary components results in the following conditions represented by equations (43) and (44).

\begin{equation}
    \sum_{k=-3}^{3} k \theta_{w_k}^R(\zeta)=0
\end{equation}

\begin{equation}
    \sum_{k=-3}^{3} k \theta_{w_k}^I(\zeta)=0
\end{equation}

Similarly substituting equations (40) and (41) into equation (39) results in the following set of conditions upon simplification:

\begin{equation}
    \sum_{k=-3}^{3} i^k k \theta_{w_k}^R(\zeta)=0
\end{equation}

\begin{equation}
    \sum_{k=-3}^{3} i^k k \theta_{w_k}^I(\zeta)=0
\end{equation}

Solving equations (41) to (44) we obtain:

\begin{equation}
    \theta_{w_3}^R(\zeta)=\frac{\theta_{w_1}^R(\zeta)}{3}
\end{equation}

\begin{equation}
    \theta_{w_3}^I(\zeta)=\frac{-\theta_{w_1}^I(\zeta)}{3}
\end{equation}

\begin{equation}
    \theta_{w_2}^I(\zeta)=0
\end{equation}

Thus, we need to employ equations (47) to (49) to account for the adiabatic boundary condition provided at the either ends of the heat exchanger wall (at $x = 0$ and $x = L$).

\subsection{Heat transfer coefficient correlation used in the 1-D Conjugate CNCL model }

In order to completely predict the transient behaviour of the Conjugate CNCL system, we need to utilise a heat transfer coefficient correlation to predict the magnitude of heat transfer coefficients on the Loop-1 and Loop-2 sides of the coupled heat exchanger. The Prandtl number of the fluid used for the 3-D CFD study is $0.000125$, so the correlations employed to predict heat transfer coefficients of liquid metals must be used in the 1-D model as they also have Prandtl numbers in the same order of magnitude. Mochizuki \cite{mochizuki2015liquid} identified the Seban and Shimazaki correlation \cite{seban1949heat} to be most appropriate for prediction the heat transfer coefficients of liquid metals. The Seban and Shimazaki correlation is as follows:

\begin{equation}
    Nu=5 + 0.025(Pe)^{0.8}
\end{equation}

The Seban and Shimazaki correlation \cite{seban1949heat} was developed for heat transfer across a circular duct, but since the heat transfer area of interest in the present study is a flat plate heat exchanger we need to employ a scaled version of equation (50), which is represented as follows:

\begin{equation}
    Nu=\frac{\pi D_h}{4L}(5 + 0.025(Pe)^{0.8})
\end{equation}

This implies that the heat transfer coefficient at the common heat exchange section is given as follows:

\begin{equation}
    U=\frac{\pi \kappa_i}{4L}(5 + 0.025(Pe)^{0.8})
\end{equation}

Thus, the Stanton number can be expressed as:

\begin{equation}
    St_i=\frac{Ut_0}{\rho_i C_{p,i} D_h}=\frac{t_0}{\rho_i C_{p,i} D_h}\times \frac{\pi \kappa_i}{4L}(5 + 0.025(Pe)^{0.8})
\end{equation}

Equation (53) is employed in the 1-D Conjugate CNCL model to predict the rate of heat exchange between the fluid and the wall in both Loop 1 and Loop 2, which helps in the prediction of the transient dynamics of the conjugate CNCL system.

\subsection{Solution of system of ODEs}

The transient non-dimensional behaviour of the conjugate CNCL system is obtained by integrating the equations (29) to (36) temporally after expanding them considering three Fourier nodes. The present study employs MATLAB ode15s solver to obtain the transient dynamics of the conjugate CNCL system. This enables us to predict the transient behaviour of every Fourier node considered.

\subsection{Fourier node independence test}

To identify the number of nodes which are required to analyse the conjugate CNCL system using the present model it is pertinent to identify the minimum number of Fourier nodes ($n$). This is accomplished by conducting a Fourier node independence test. Figure 2 depicts the Fourier node independence test for the case CCNCL-(c) (described in the next section). The variables $Re$ and $\theta_{Avg}$ of Loop 1 of the conjugate CNCL system are utilised to carry out the Fourier node independence test. The case CCNCL-(c) is used as it has the maximum magnitude of $Re$ and $\theta_{Avg}$. From Figure 2 it is evident that n=3 is adequate to fully describe the transient characteristics of the conjugate CNCL system.

\begin{figure}[!htb]
    \centering
    \includegraphics[width=0.6\linewidth]{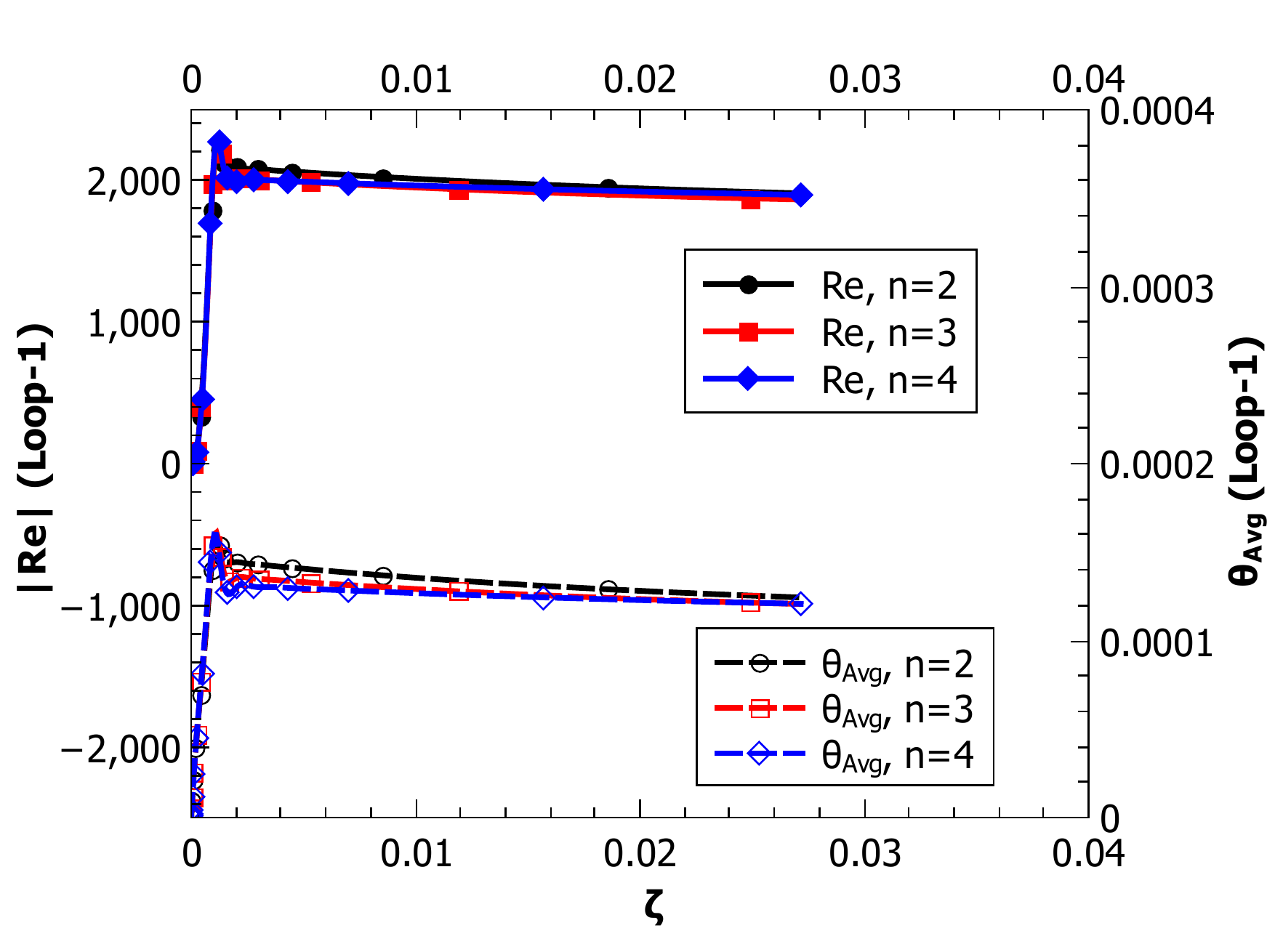}
    \caption{Fourier node independence test of the 1-D Conjugate CNCL model.}
    \label{FourierNodeIndependenceTest}
\end{figure}

\section{3-D CFD study of conjugate CNCL system}

To verify the 1-D model of the conjugate CNCL system a systematic 3-D CFD study is undertaken with ANSYS Fluent 16.1 software. Table 1 lists the CFD cases studied in the present work. Four CFD cases are studied to determine the influence of wall thickness and inclination on the conjugate CNCL system. The case CCNCL-(a) of Table 1 represents the 3-D CFD case with $0\;mm$ wall thickness and zero inclination.

\begin{table}[!htb]
\centering
\caption{Cases considered for the 3-D CFD study.}
\begin{tabular}{|c|c|c|}
\hline
\textbf{Case}      & \textbf{Heat exchanger wall thickness ($t_w$)} & \textbf{Conjugate CNCL inclination w.r.t gravity ($\phi$)} \\ \hline
\textbf{CCNCL-(a)} & $0\; mm$                                       & $0^\circ$                                                  \\
\textbf{CCNCL-(b)} & $1\; mm$                                       & $0^\circ$                                                  \\
\textbf{CCNCL-(c)} & $5\; mm$                                       & $0^\circ$                                                  \\
\textbf{CCNCL-(d)} & $5\; mm$                                       & $45^\circ$       \\ \hline                                         
\end{tabular}
\end{table}

\subsection{Pre-processing stage}

\subsubsection{Geometry of the Conjugate CNCL system}

The geometry of the Conjugate CNCL system is constructed by considering two individual Natural Circulation loops (NCL) which are connected to each other via the common wall of thickness $t_w$. The considered NCLs have a square cross section of length $D_h$ with height $L$ and width $L1$. Figure 3 depicts the schematic of the geometry used for the current study. Table 2 lists the dimensions of the Conjugate CNCL used for the present CFD study.

\begin{table}[!htb]
\centering
\caption{Dimensions of the Conjugate CNCL system used for 3-D CFD study.}
\begin{tabular}{|l|l|l|}
\hline
\textbf{Parameter} & \textbf{Description}     & \textbf{Magnitude}  \\ \hline
$L$ & Height of the conjugate CNCL & $1\; m$                                                                                         \\
$L1$ & Width of Loop 1 and Loop 2 &  $1\; m$                                                                                         \\
$D_h$ & Length of square cross section &$0.04\; m$                                                                                         \\
$t_w$ & Thickness of the common heat exchanger wall & $0-5\; mm$                                              \\ \hline                                         
\end{tabular}
\end{table}

\begin{figure}[!htb]
    \centering
    \includegraphics[width=0.7\linewidth]{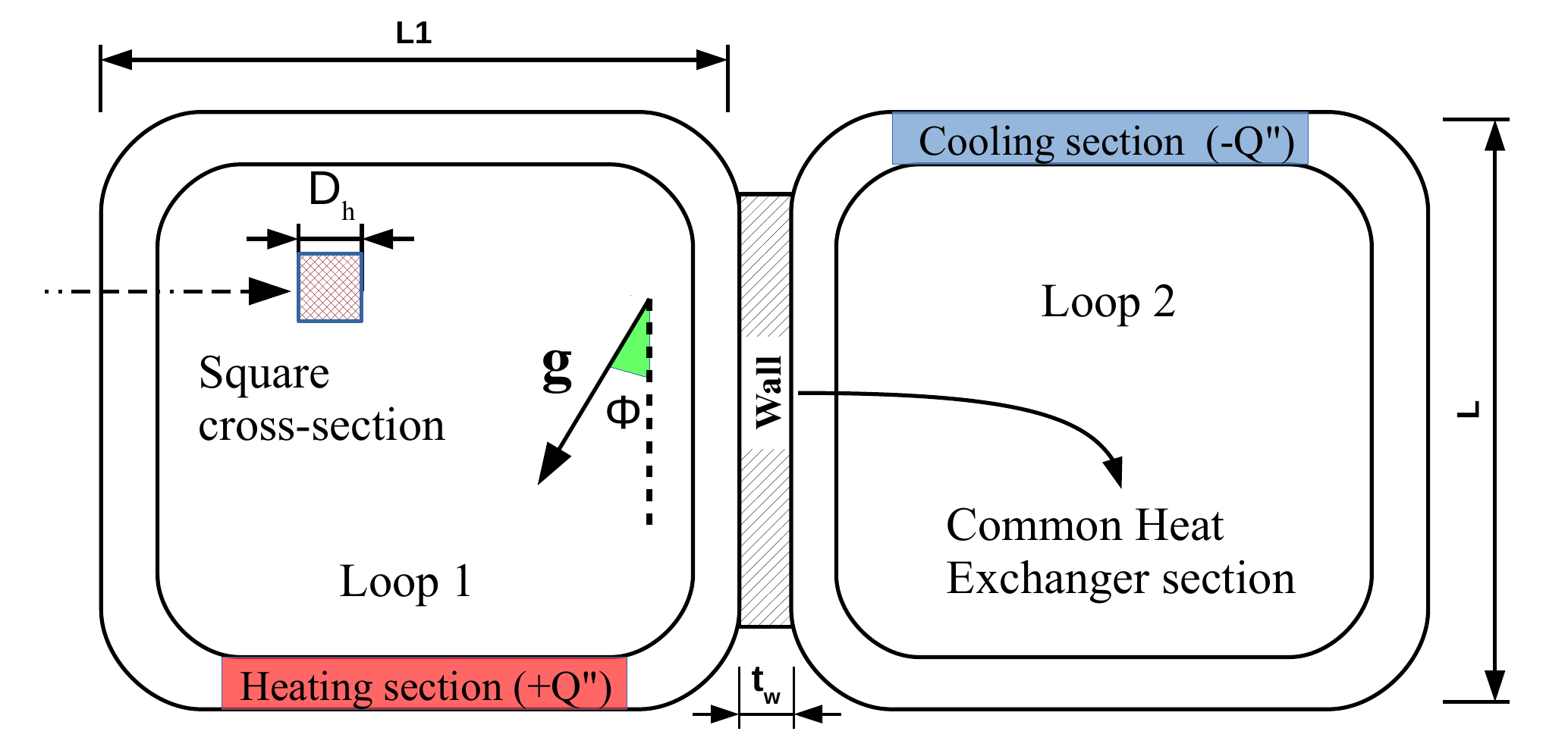}
    \caption{Schematic of the geometry used for the 3-D CFD study of the Conjugate CNCL system.}
    \label{3DCFDGeometry}
\end{figure}

\subsubsection{Meshing the Conjugate CNCL system}

The geometry is meshed using the ANSYS Meshing software. A structured mesh was generated using the multizone method with an average element quality of $0.98$. The schematic of the mesh used for the present study is shown in Figure 4.

\begin{figure}[!htb]
    \centering
    \includegraphics[width=\linewidth]{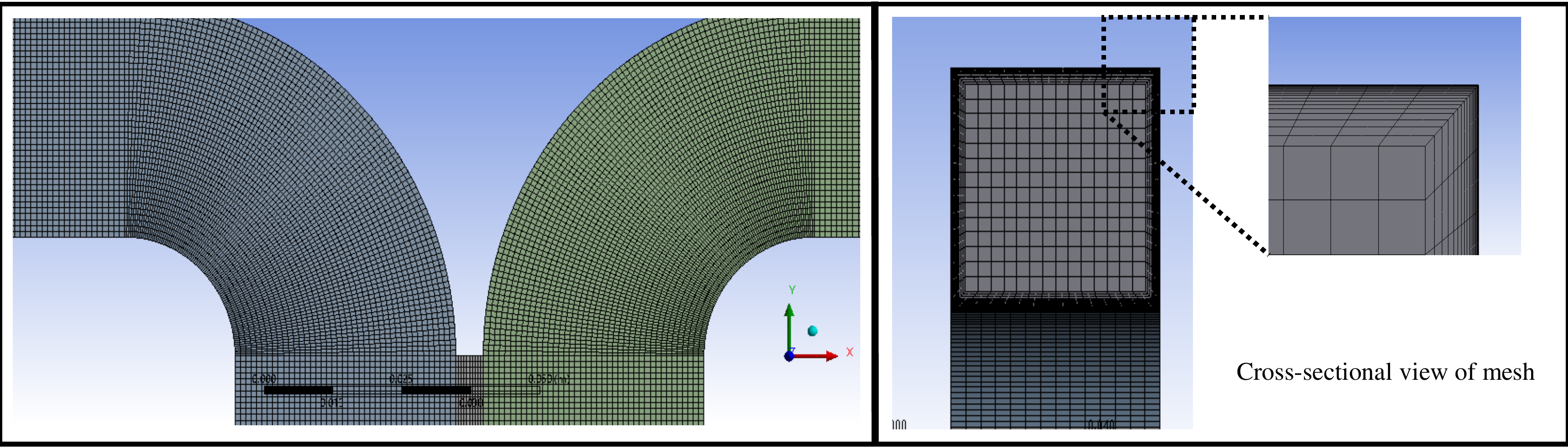}
    \caption{Schematic of the mesh used for the 3-D CFD study of the Conjugate CNCL system.}
    \label{3DCFDMesh}
\end{figure}

\subsection{Processing stage}

\subsubsection{Case setup}

A transient pressure-based solver is utilised without any turbulence model as the flow is studied within the laminar flow regime. The Boussinesq approximation is used to model the dependence of fluid density on temperature. The only input given to the system is the constant heat flux boundary condition at the heating and cooling sections respectively. The entire system is at an initial temperature of $T_0$ and zero initial velocity. A PISO scheme with second order upwind for momentum and energy discretization is used. A transient second order implicit time scheme has been used for the present 3-D study. The present study employs a fictitious fluid to reduce the computational load. Fictitious fluids were also used by Dass and Gedupudi (2019) \cite{dass2019} to speed up the transient flow evolution and relax the grid and time step size requirements for quicker computation. The magnitude of the thermophysical properties of the fluid and the wall are listed in Table 3 and the values of the initial and boundary conditions used for the CFD study are listed in Table 4. The dependence of the fluid and wall properties can be eliminated by non-dimensionalisation. Thus, even if we employ different materials which have different thermophysical properties but have the same magnitudes of non-dimensional terms then the non-dimensional flow characterization is independent of the fluid and wall properties.The magnitudes of the non-dimensional numbers used for the present CFD study are listed in Table 5. Only the magnitude of $Co_B$ changes for all the CFD cases as it is a function of $t_w$. 

\begin{table}[h!]
	\centering
	\caption{Thermophysical properties of the materials used in the 3-D CFD study.}
	\label{Table3}
	\scalebox{1}{
		\begin{tabular}{  |c | c | c | }
			\hline
			\textbf{Thermophysical property}              & \textbf{Fluid}  & \textbf{Wall}    \\ \hline
			$\rho_0 $ $(kg/m^3)$  &      50          &  8978    \\
			$C_p$ $(J/kgK)$      &      5         &   381    \\
			$\beta$ $(1/K)$      &      0.08            &   -   \\
			$\alpha$ $(m^2/s)$          &      0.8           &   0.00011
  \\
			$\mu$ $(kg/ms)$        &    0.005         &  - \\ \hline
	\end{tabular}}
\end{table}

\begin{table}[h!]
	\centering
	\caption{Initial and boundary conditions}
	\label{Table4}
	\scalebox{1}{
		\begin{tabular}{ | c | c | c | }
			\hline
			\textbf{Parameter}     & \textbf{Description}  & \textbf{Values}    \\ \hline
			$T_{0} $      & Temperature at time $t=0$ &  300 $K$    \\
			$P_0$         & Pressure at time $t=0$    &   1 $atm$    \\
			$Q"$          & Constant heat flux supplied or extracted &  2000 $W/m^2$    \\
			$g$           & Gravitational constant    &  9.81 $m/s^2$   \\ \hline
	\end{tabular}}
\end{table}

\begin{table}[!htb]
\centering
\caption{Magnitude of the non-dimensional numbers representing the considered 3-D CFD cases.}
\begin{tabular}{|l|l|l|l|l|l|l|}
\hline
\textbf{Case}      & \textbf{$Co_B$} & \textbf{$Gr$}                         & \textbf{$Fo$}              & \textbf{$Fo_w$}           & \textbf{$Pr$}               & \textbf{$Co_A$}   \\  \hline       
\textbf{CCNCL-(a)} & 0.2923          & \multirow{4}{*}{$6.8 \times 10^{10}$} & \multirow{4}{*}{$173.913$} & \multirow{4}{*}{$0.0246$} & \multirow{4}{*}{$0.000125$} & \multirow{4}{*}{$1309.16$} \\
\textbf{CCNCL-(b)} & 0.002923        &                                       &                            &                           &                             &                            \\
\textbf{CCNCL-(c)} & 0.000585        &                                       &                            &                           &                             &                            \\
\textbf{CCNCL-(d)} & 0.000585        &                                       &                            &                           &                             &      \\   \hline                  
\end{tabular}
\end{table}

\subsubsection{Grid and time step independence test}

To ascertain the accuracy of the CFD study, a grid and time step independence study has been conducted. The number of elements of the grid was increased and the time step used was decreased till the change in the transient behaviour was independent of both. Figure 5a and Figure 5b present the grid and time step study evaluated for the present study. It is observed that a mesh with number of elements of 10 lakhs and time step size of $0.2s$ were adequate for the transient study of the case CCNCL-(c) which corresponds to $t_w=5 mm$. Case CCNCL-(c) has the maximum Reynolds number and non-dimensional fluid averaged temperature for the considered CFD study as will be noted in the upcoming sections. Thus, it can be safely concluded that the above considered number of elements and time step size are adequate for all the CFD cases considered.

\begin{figure}[!htb]
	\centering
	\begin{subfigure}[b]{0.49\textwidth}
		\includegraphics[width=1\linewidth]{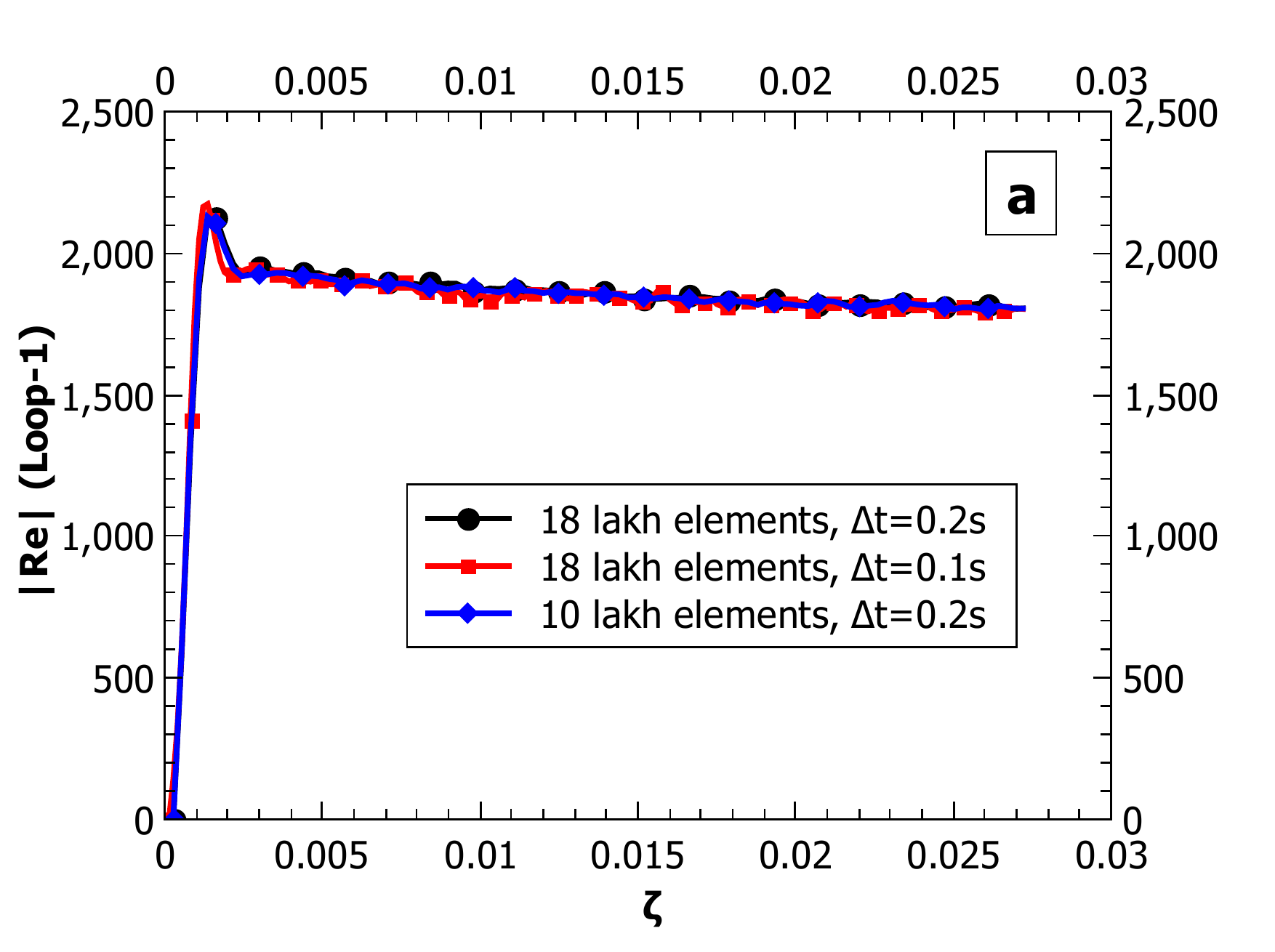}
	\end{subfigure}
	\hspace{\fill}
	\begin{subfigure}[b]{0.49\textwidth}
		\includegraphics[width=1\linewidth]{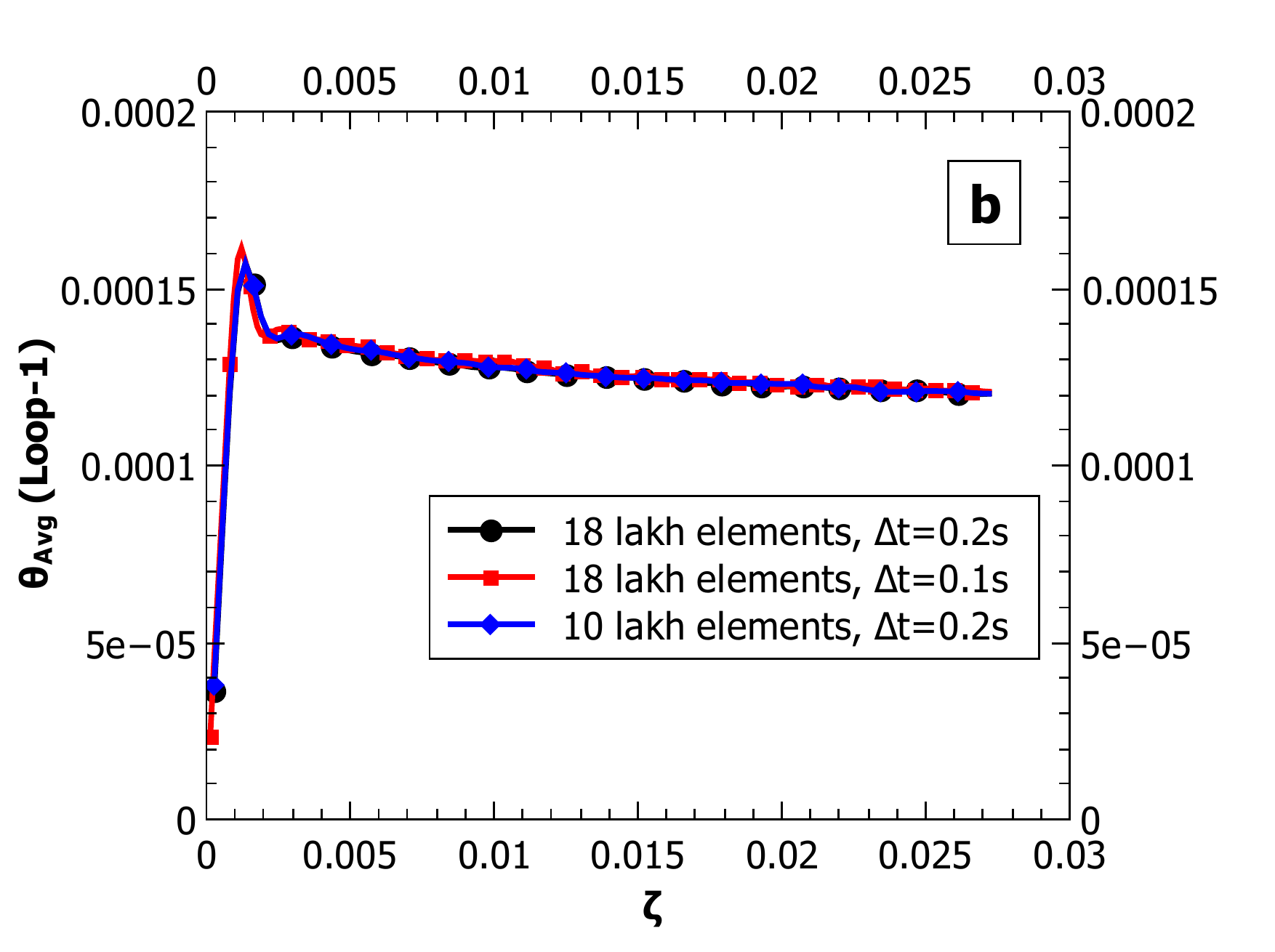}
	\end{subfigure}
	\caption{Grid and time step independence study conducted using (a) Reynolds number ($Re$) as the flow parameter (b)  Non-dimensional fluid average temperature ($\theta_{Avg}$) of Loop 1 as the flow parameter. }
	\label{Mesh and timestep independence test}
\end{figure}

\subsubsection{Validation of CFD methodology}

\begin{figure}[!htb]
    \centering
    \includegraphics[width=0.5\linewidth]{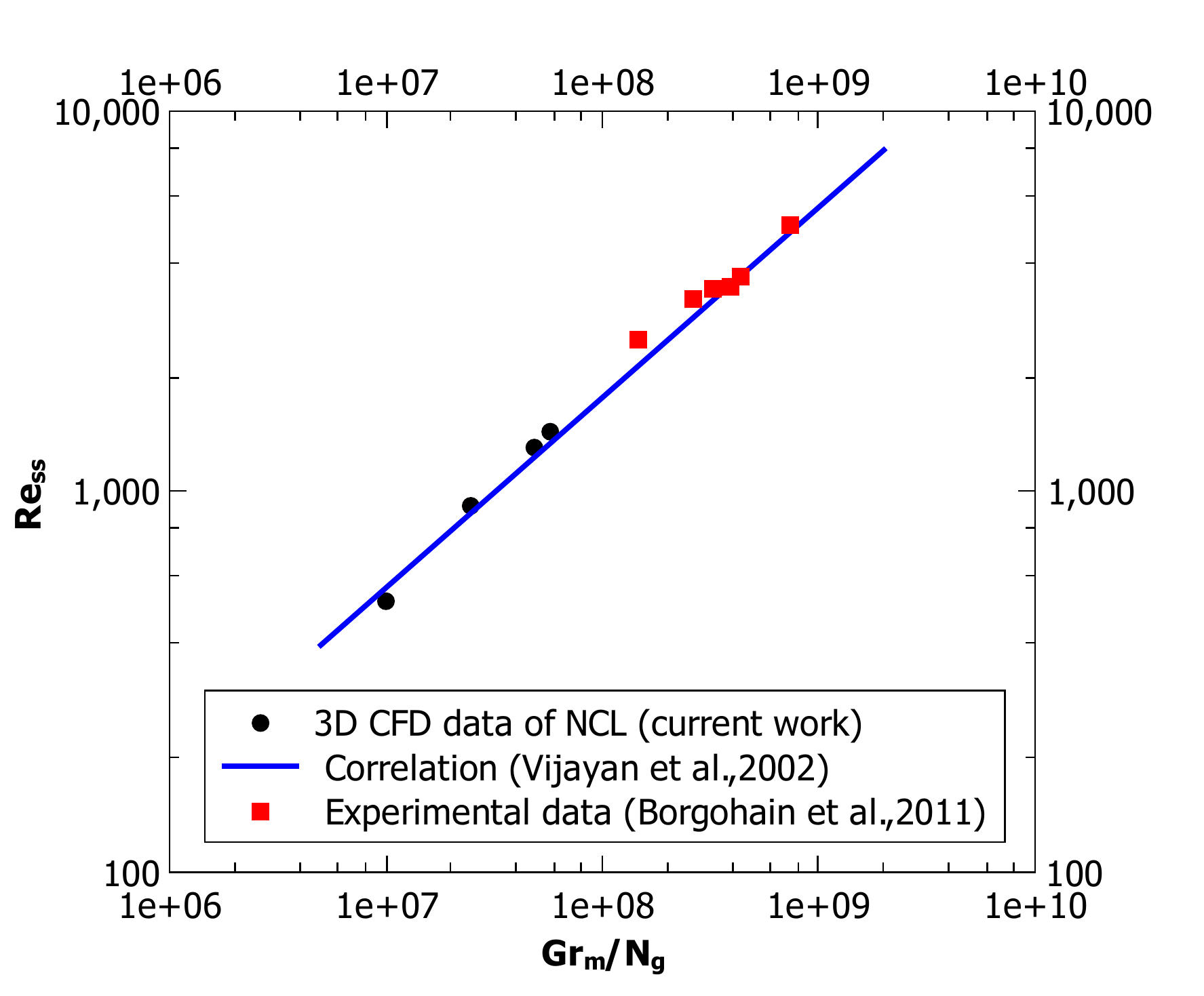}
    \caption{Validation study of the methodology employed for the present 3-D CFD study. }
    \label{3DCFDmethodologyValidation}
\end{figure}

It is necessary to validate the CFD methodology to ensure that the schemes used for the CFD study are capable of capturing the accurate physics of the phenomena. This can be ensured by comparing the CFD prediction with experimental data. Since there is no direct experimental data available on CNCL systems to validate the CFD methodology we utilise the experimental correlation by Vijayan et al. \cite{vijayan2002} for single phase NCLs and compare it with the data obtained from the 3-D CFD study of an equivalent NCL with similar boundary conditions. Figure 6 represents the comparison between the experimental correlation and the CFD prediction and a good match is observed indicating the suitability of the settings utilised for the CFD study. The experimental correlation of Vijayan et al. \cite{vijayan2002} shows a good match with the experimental data of Borgohain et al. \cite{borgohain2011} indicating the appropriateness of the correlation for describing the natural circulation phenomena. The correlation developed by Vijayan et al. \cite{vijayan2002} for natural circulation flows is as follows:

\begin{equation}
    Re_{ss}=0.1768(\frac{Gr_m}{N_g})^{0.5}
\end{equation}

\subsection{Post-processing stage}

\subsubsection{Results from the 3-D CFD study}

\begin{figure}[!htb]
	\centering
	\begin{subfigure}[b]{0.8\textwidth}
		\includegraphics[width=1\linewidth]{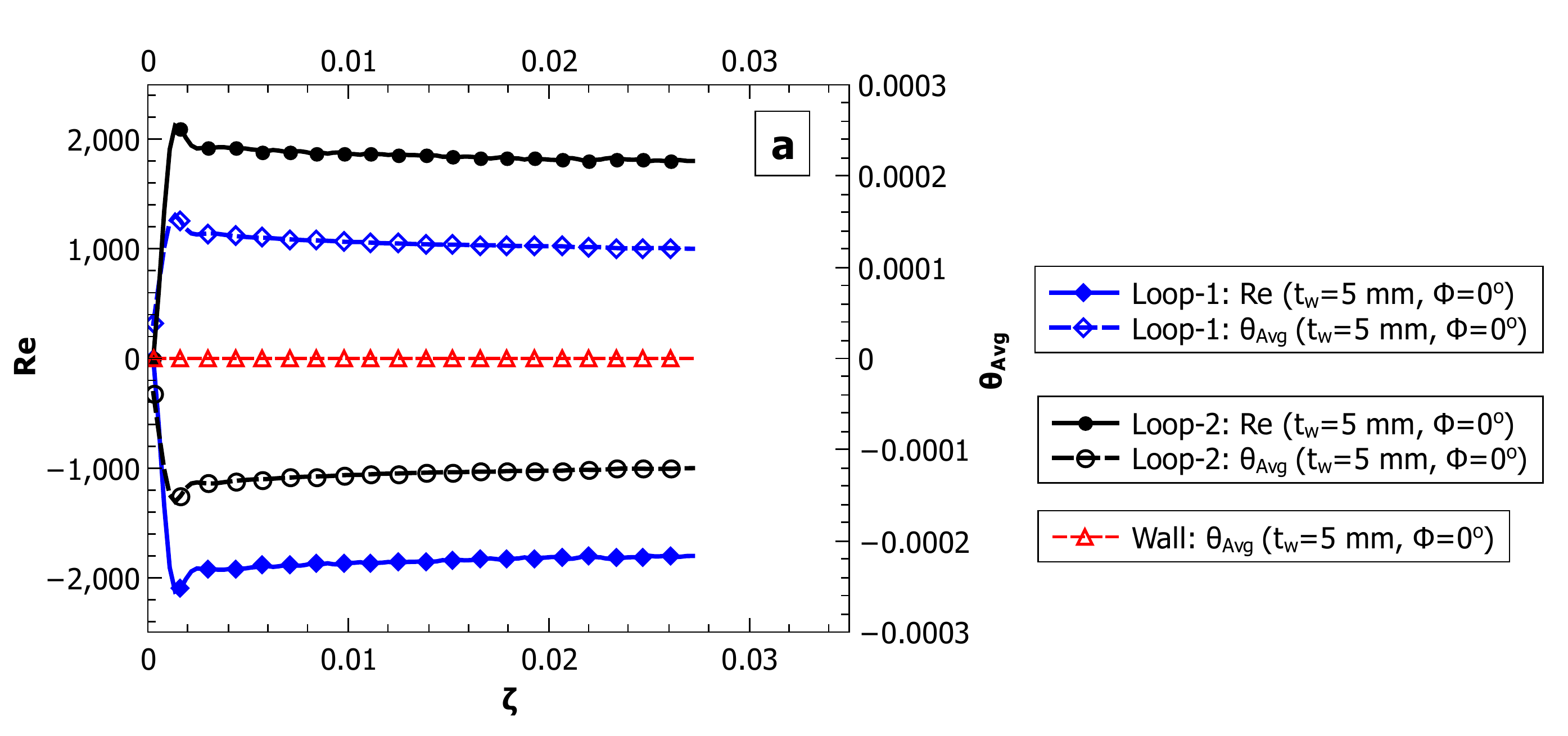}
	\end{subfigure}
	\hspace{\fill}
	\begin{subfigure}[b]{0.8\textwidth}
		\includegraphics[width=1\linewidth]{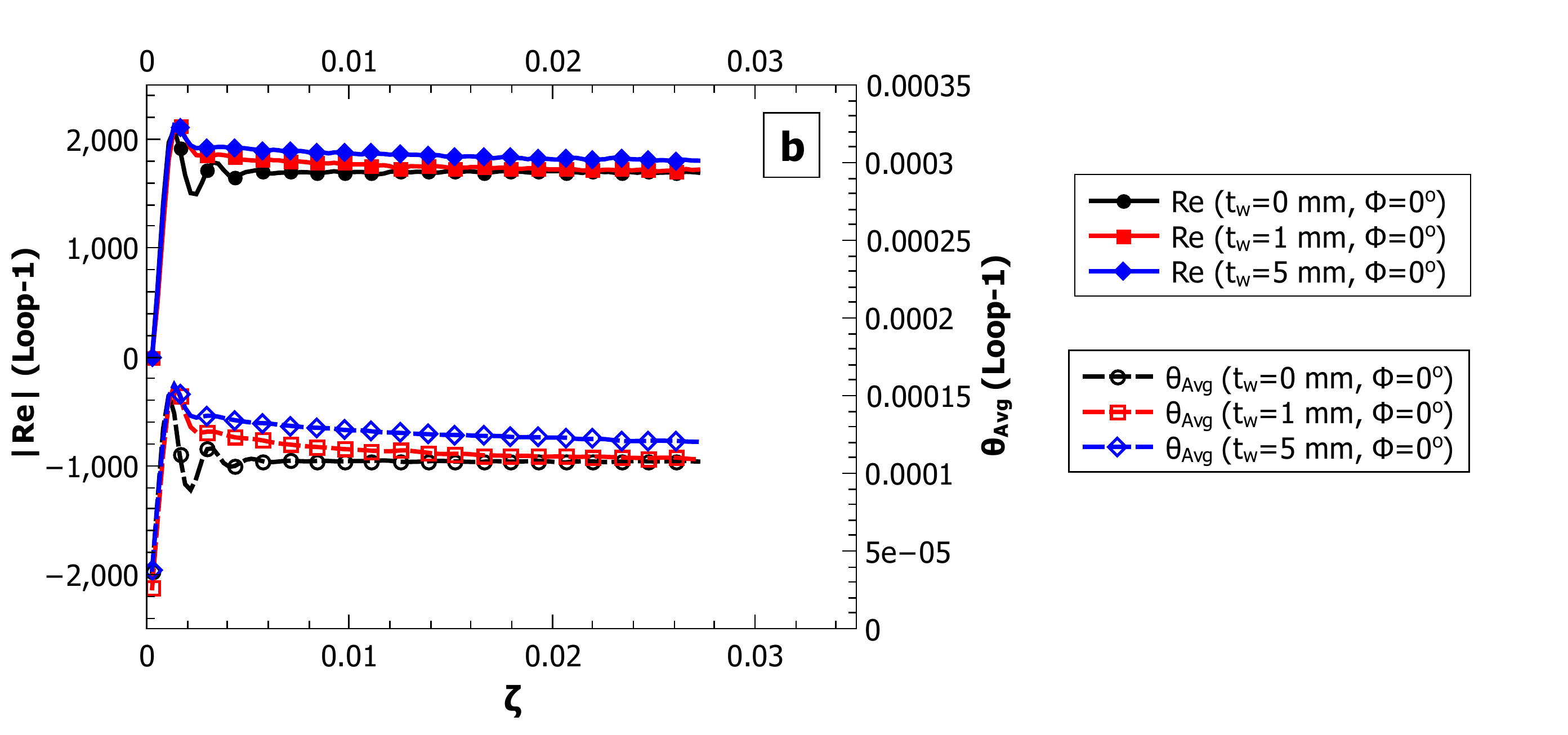}
	\end{subfigure}
	\caption{(a) Transient temperature variation of $Re$ and $\theta_{Avg}$ of Loop-1 and Loop-2 of the 3-D CFD case CCNCL-(c) (b) Effect of wall thickness on the transient behaviour of $Re$ and $\theta_{Avg}$ of Loop-1 from 3-D CFD study.}
	\label{3DCFDResults_a}
\end{figure}

Figure 7 represents the 3-D CFD results of the conjugate CNCL system with $\phi=0^{\circ}$. It is to be noted that the parameter $Re$ is assigned a sign consistent with the developed 1-D model to denote the flow direction. A ‘-Ve’ sign of $Re$ for Loop-1 indicates that the flow has a clockwise direction in Loop-1 from the readers perspective and a ‘+Ve’ sign indicates the opposite flow direction. A ‘+Ve’ sign of $Re$ for Loop-2 signifies an clockwise flow within Loop-2 from the readers perspective and a ‘-Ve’ sign implies anti-clockwise flow direction. Figure 7a presents the transient behaviour of parameters $Re$ and $\theta_{Avg}$ of Loop-1, Loop-2 and $\theta_{Avg}$ of Wall. A symmetric transient behaviour of $Re$ and $\theta_{Avg}$ of Loop-1 and Loop-2 is observed with respect to $\theta_{Avg}$ of the Wall. This is due to the symmetry of the system and the chosen heater cooler configuration. All the considered cases in the present study exhibit similar behaviour. As a result of this symmetry, it is adequate to present the results of Loop-1. The symmetry of the system also leads to the magnitude of $\theta_{Avg}$ being zero at all instances of time for all the considered CFD cases and is thus not represented specifically in the graphs in the remaining part of this paper.

Figure 7b shows the effect of wall thickness on the transient behaviour of the natural circulation flow in each loop. We observe that the wall thickness has a significant influence on the transient trend of the conjugate CNCL system. The Reynolds number and non-dimensional fluid averaged temperature increase in magnitude with increase in the thickness of the heat exchanger wall. The initial transient oscillatory behaviour observed for the no wall case is not observed with the inclusion of the heat exchanger wall thickness ($t_w$).

\begin{figure}[!htb]
    \centering
    \includegraphics[width=0.7\linewidth]{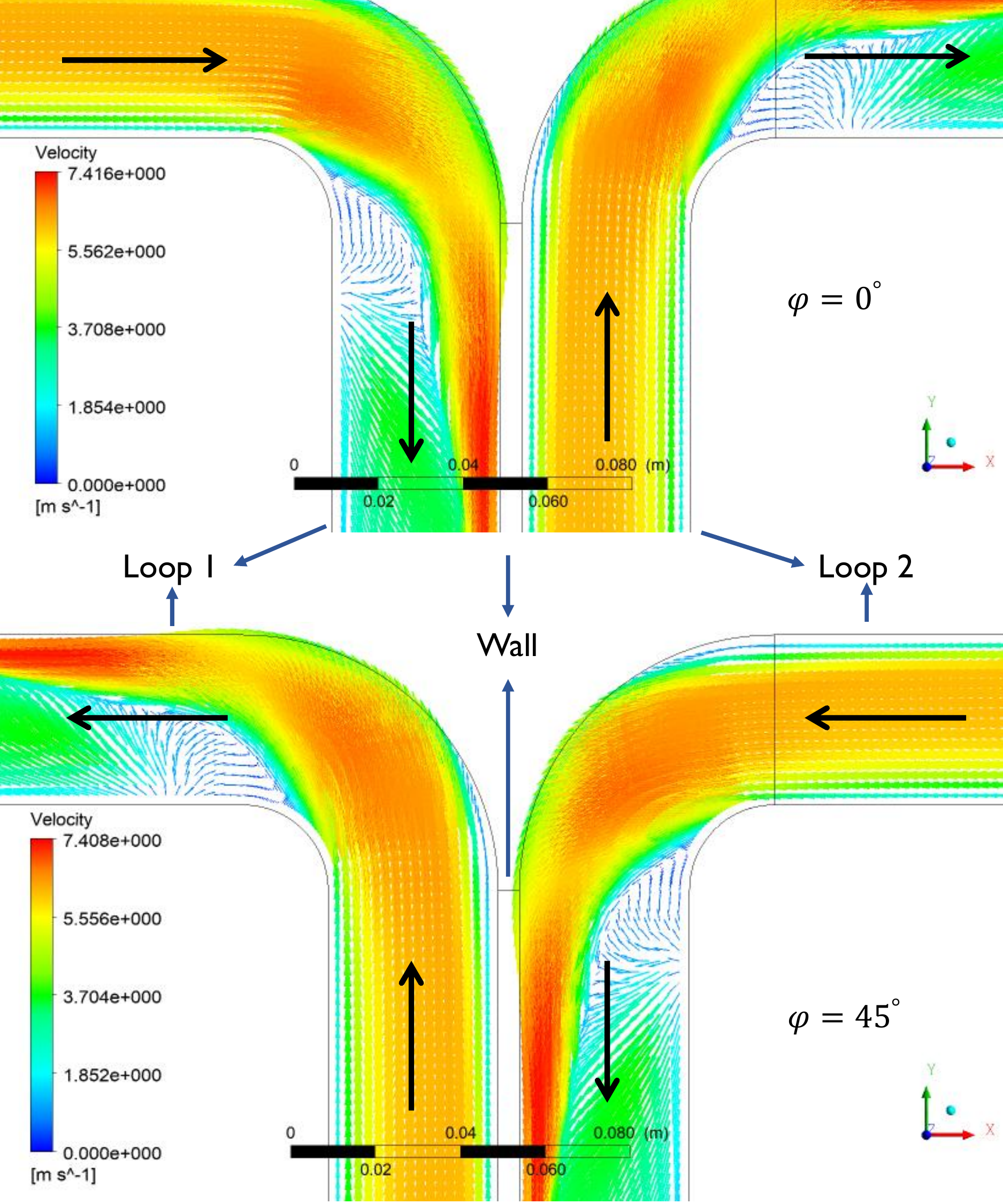}
    \caption{Velocity vector plot of the mid-plane of 3-D CFD case at $\phi=0^{\circ}$ corresponding to case CCNCL-(c) and $\phi=45^{\circ}$ corresponding to case CCNCL-(d). The black arrows indicate the direction of flow in each of the loops of the system.}
    \label{3DCFDResults_b}
\end{figure}

Figure 8 represents the mid-plane velocity contours of CFD cases CCNCL-(c) and CCNCL-(d) where the wall thickness is constant ($t_w=5\; mm$) and the inclination is varied from $\phi=0^{\circ}$ to $\phi=45^{\circ}$. It is observed that there is a flow direction reversal in both Loop-1 and Loop-2 with change in inclination for the considered cases.

\subsection{Verification of the 1-D semi-analytical model of the Conjugate CNCL system}

This section presents a detailed verification of the developed 1-D model for the inclined conjugate CNCL system with rectangular loops. To verify the model extensively, it is compared to the data available in the existing literature and with the 3-D CFD cases conducted in the present work. 

\subsubsection{Verification of the 1-D model with existing literature on CNCL systems}

Dass and Gedupudi \cite{dass2019} conducted a transient 3-D CFD study of a single phase rectangular CNCL system with similar heater-cooler configuration for a no wall and zero inclination case. A different fluid was used for their study. Thus, the CNCL system for their study can be used as a reference for verification of the 1-D model of the inclined conjugate CNCL model. Figure 9 represents the comparison of the data from Dass and Gedupudi \cite{dass2019} with the developed 1-D model and a good agreement is observed in the transient trends of $Re$ and $\theta_{Avg}$, which demonstrates the accuracy of the developed model. It also demonstrates that the model works irrespective of the fluid used as long as the Seban and Shimazaki correlation \cite{seban1949heat} can be used to predict the heat transfer. This completes the verification of the 1-D model with available literature data on CNCL systems. 

\begin{figure}[!htb]
    \centering
    \includegraphics[width=0.55\linewidth]{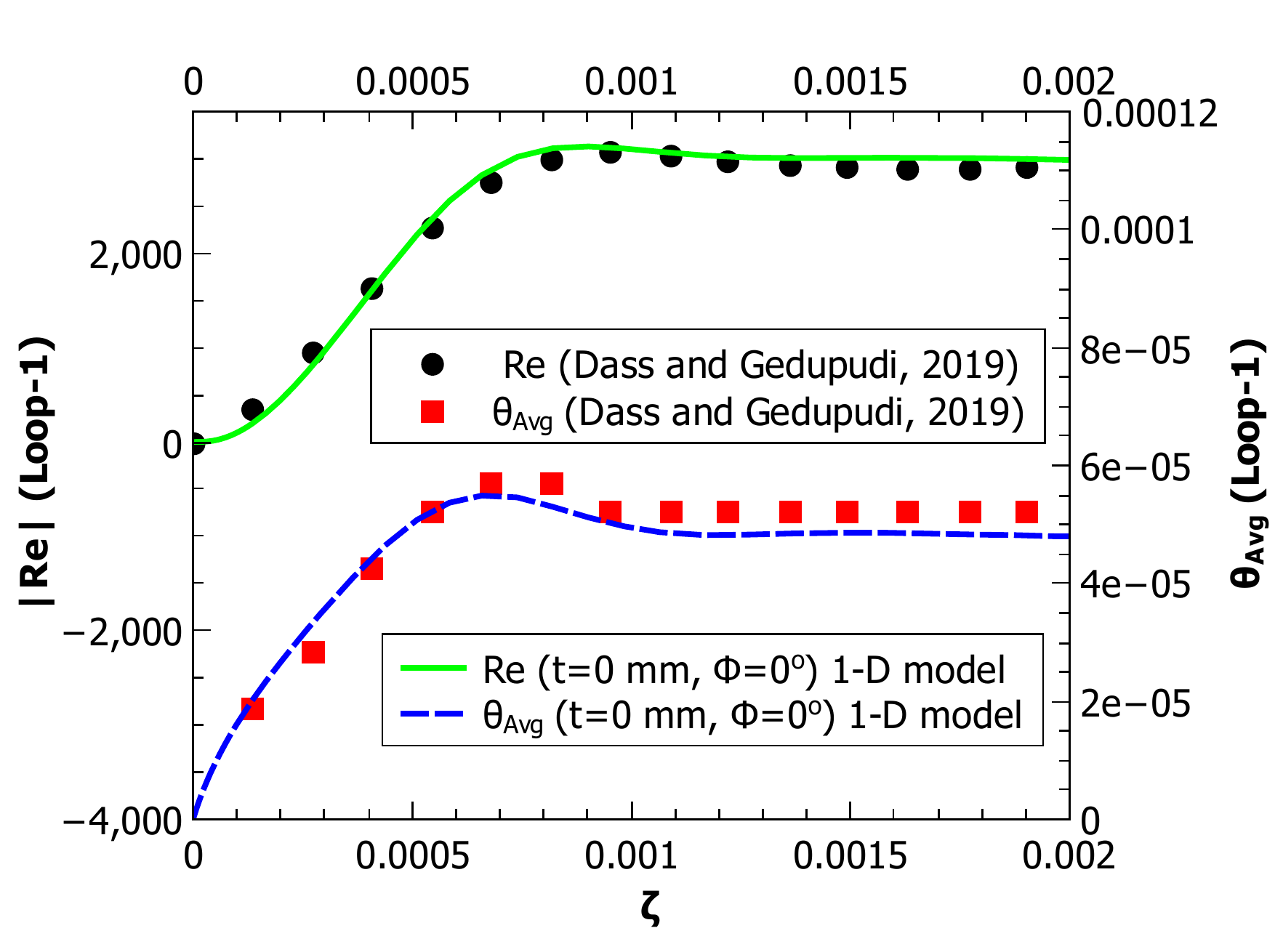}
    \caption{Verification of the 1-D model with data from Dass and Gedupudi (2019) \cite{dass2019} which corresponds to $t_w=0$ mm and $\phi=0^{\circ}$.}
    \label{Verification with existing literature}
\end{figure}

\subsubsection{Verification of the 1-D model with 3-D CFD results}

\begin{figure}[!htb]
	\centering
	\begin{subfigure}[b]{0.49\textwidth}
		\includegraphics[width=1\linewidth]{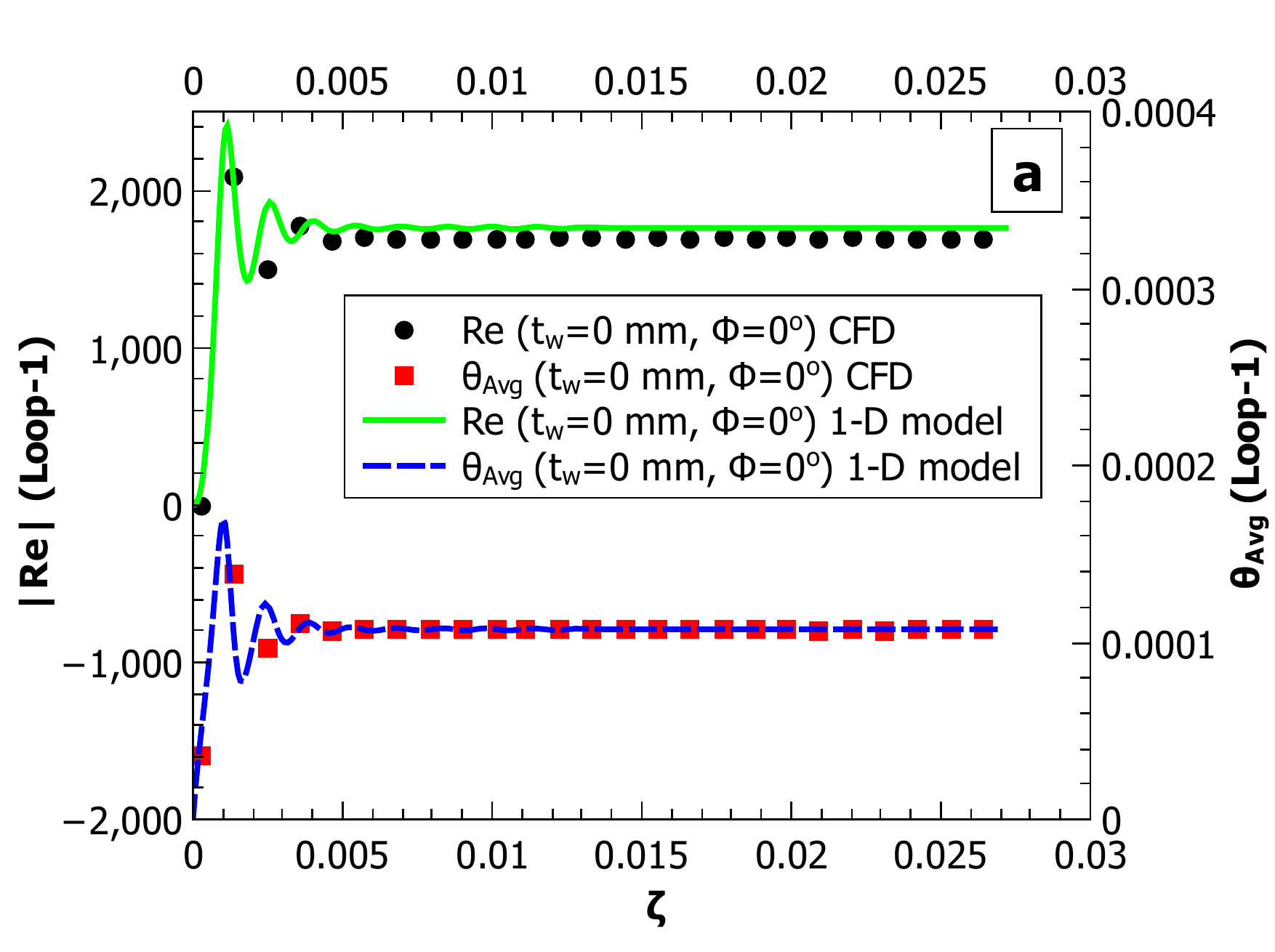}
	\end{subfigure}
	\hspace{\fill}
	\begin{subfigure}[b]{0.49\textwidth}
		\includegraphics[width=1\linewidth]{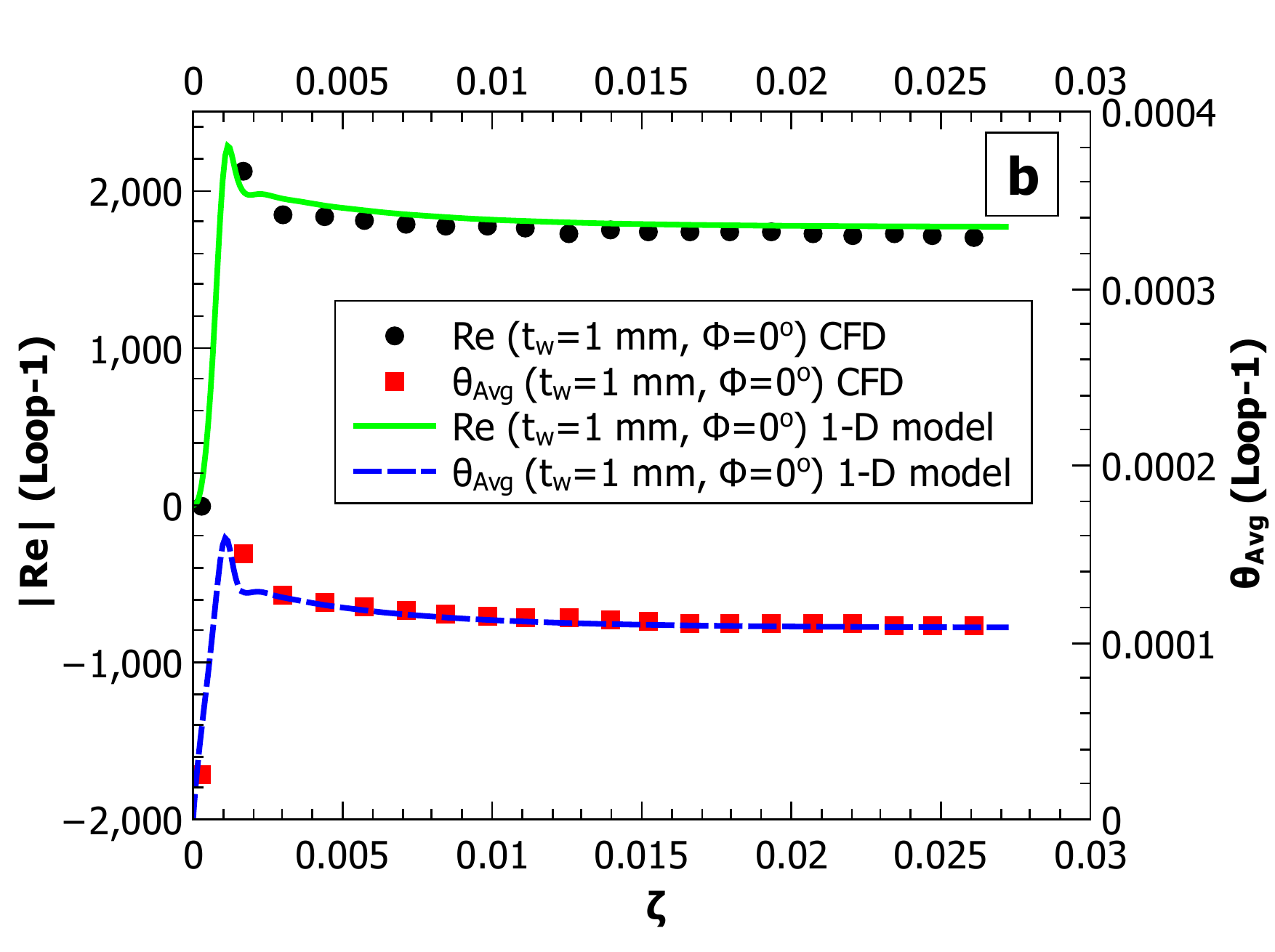}
	\end{subfigure}
	\hspace{\fill}
	\begin{subfigure}[b]{0.49\textwidth}
		\includegraphics[width=1\linewidth]{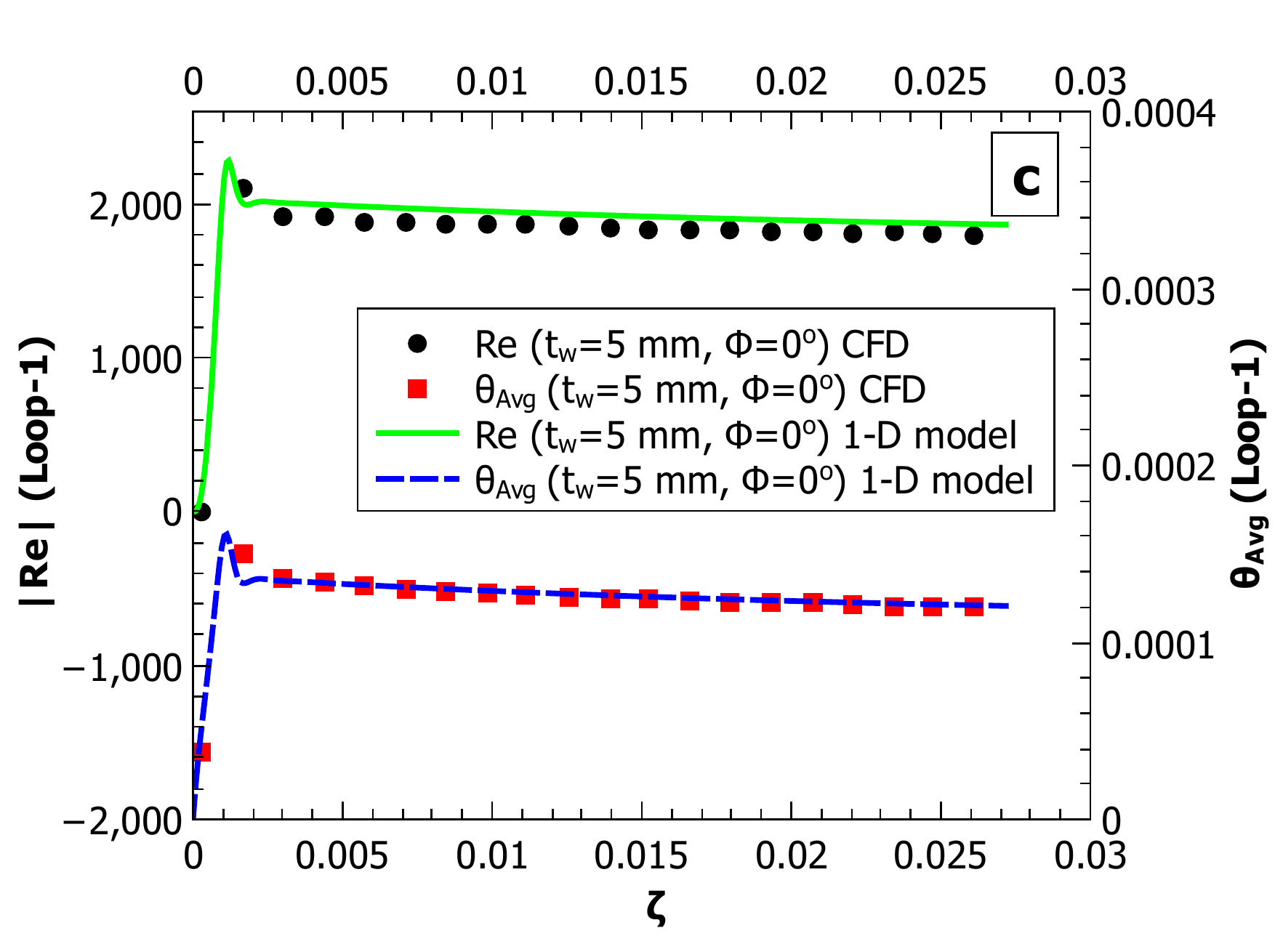}
	\end{subfigure}
	\hspace{\fill}
	\begin{subfigure}[b]{0.49\textwidth}
		\includegraphics[width=1\linewidth]{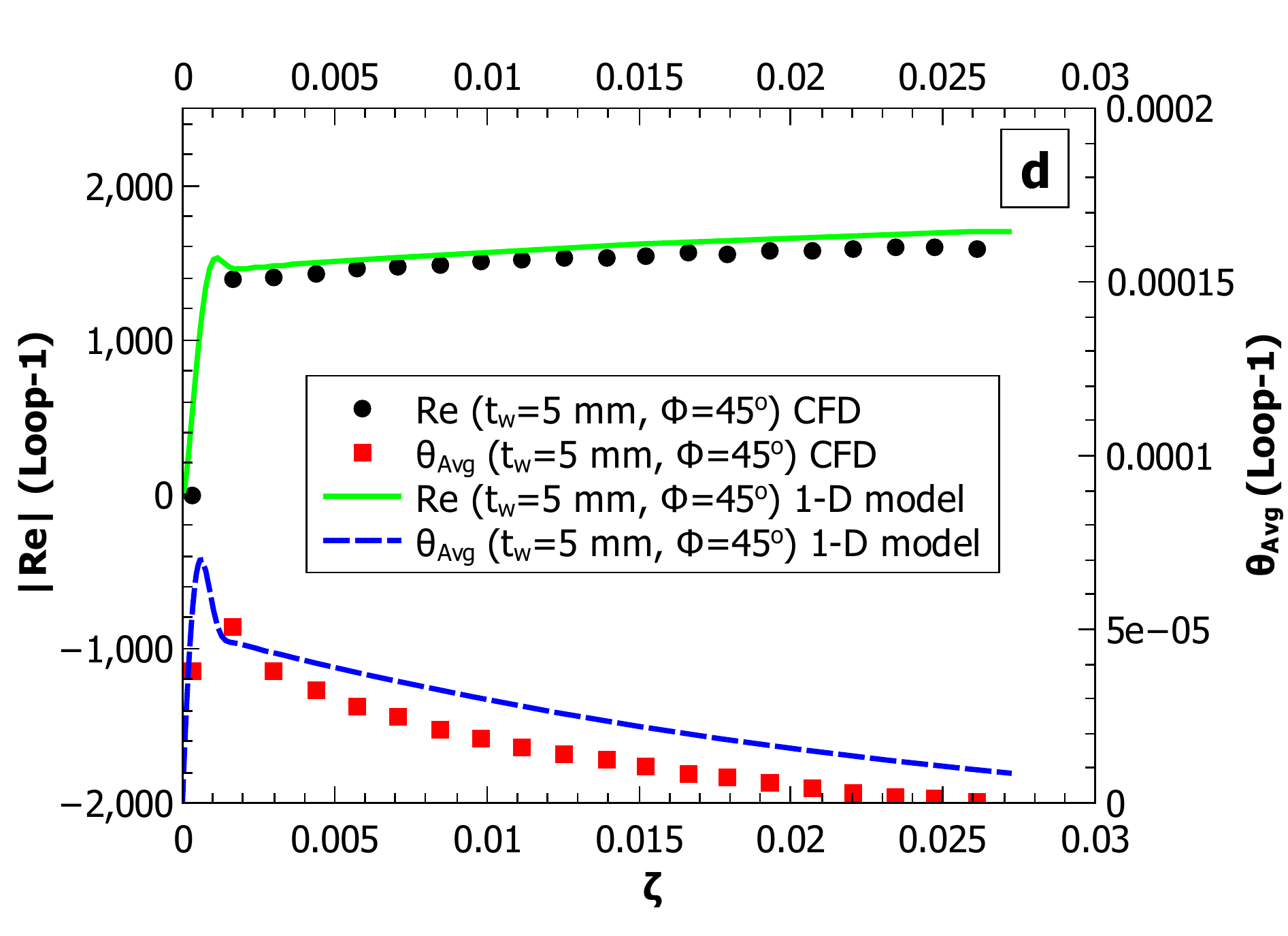}
	\end{subfigure}
	\caption{Verification of 1-D model with 3-D CFD data for (a) case CCNCL-(a) (b) case CCNCL-(b) (c) case CCNCL-(c) (d) case CCNCL-(d). }
	\label{1-D Conjugate CNCL model validation}
\end{figure}

To verify the capability of the developed 1-D semi-analytical model of the conjugate CNCL system to capture the transient dynamics of the system behaviour, a detailed verification is performed by comparing it against all the CFD cases conducted in the present paper. $L=L1=1-2R$ is used in the 1-D model, neglecting buoyancy contributions at the bends ($90^{\circ}$ smooth bends with $R/D_h=1$). Figures 10(a) to 10(d) represent the verification of the 1-D model of the conjugate CNCL against 3-D CFD cases CCNCL-(a) to CCNCL-(d). A good match between the 1-D model prediction and the 3-D CFD study is observed indicating the accuracy of the developed model for cases CCNCL-(a) to CCNCL-(c). For the case of CCNCL-(d) there is deviation of the predicted values of $\theta_{Avg}$, which may be because of the influence of inclination. This is discussed in the next section. The maximum percentage deviation between the 1-D model and 3-D CFD results is about $5\%$ for $Re$ for all the cases and the maximum percentage deviation for $\theta_{Avg}$ for cases CCNCL-(a) to CCNCL-(c) is about $15\%$. The large percentage difference is due to the magnitude of $\theta_{Avg}$ being of the order of $10^{-4}$ and because of employing the Seban and Shimazaki correlation \cite{seban1949heat} which only provides the Nusselt number magnitude at steady state for the entire transient simulation. 

\subsection{Effect of inclination on the Conjugate CNCL system}

The influence of inclination on NCL and CNCL systems has been studied by Dass and Gedupudi \cite{dass2020} employing 2-D CFD studies. They have identified that with change in inclination of the system a flow direction reversal occurs which is accompanied by a jump in the heat transfer coefficient. But their study was limited to the study of inclination systems with hysteresis effects. Since the present work is focused on the transient evolution of systems with zero flow initial condition, the authors carried out a 2-D CFD study of the CNCL, as described by Dass and Gedupudi \cite{dass2020}, but employing the zero flow initial condition in the current work. The jump in heat transfer coefficient was also observed for the case of zero initial flow condition caused by flow direction reversal. The results of the 2-D CNCL study of the CNCL with hysteresis and zero flow initial condition is presented in Figure 11.

\begin{figure}[!htb]
	\centering
	\begin{subfigure}[b]{0.6\textwidth}
		\includegraphics[width=1\linewidth]{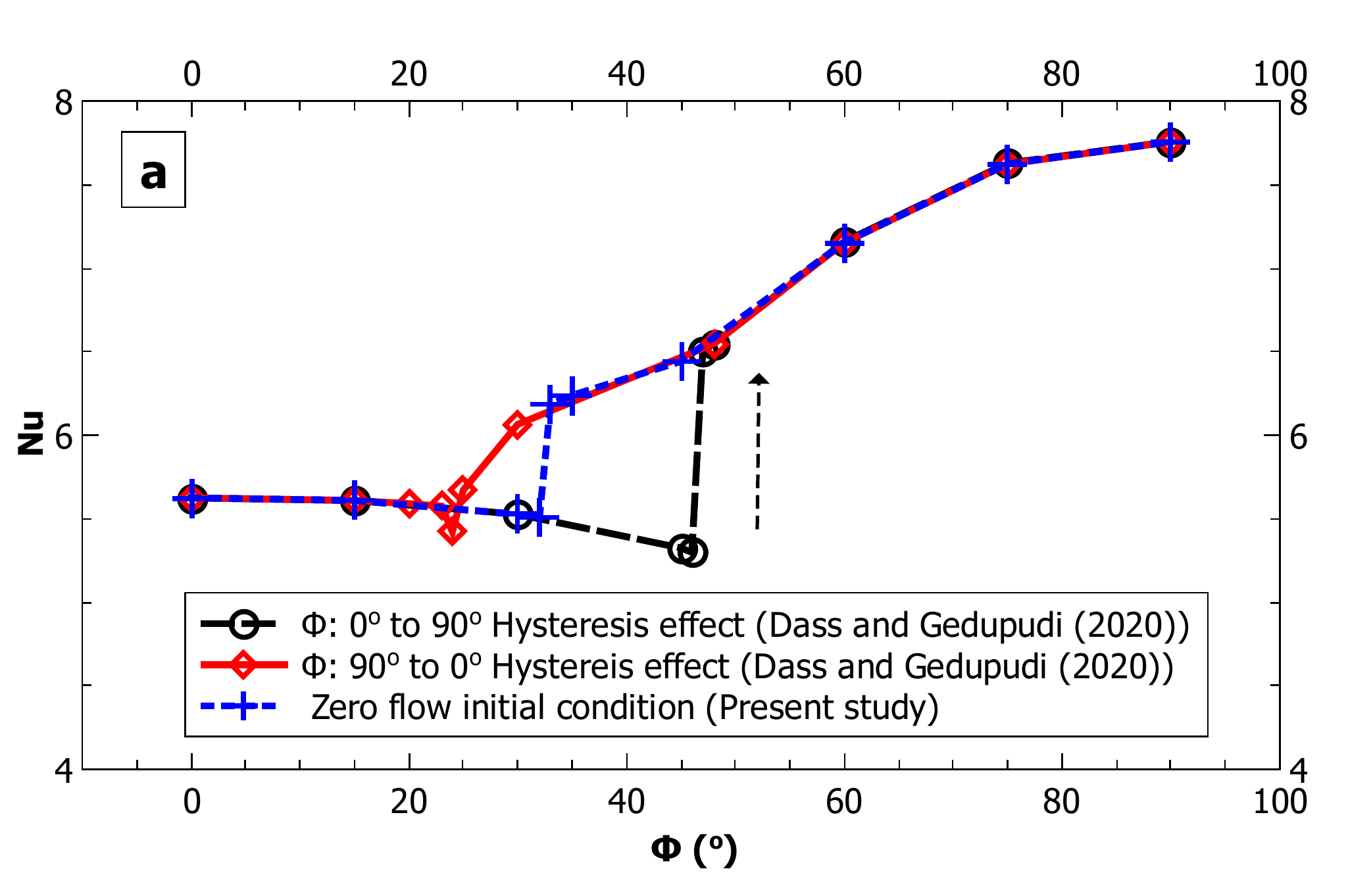}
	\end{subfigure}
	\hspace{\fill}
	\begin{subfigure}[b]{0.6\textwidth}
		\includegraphics[width=1.05\linewidth]{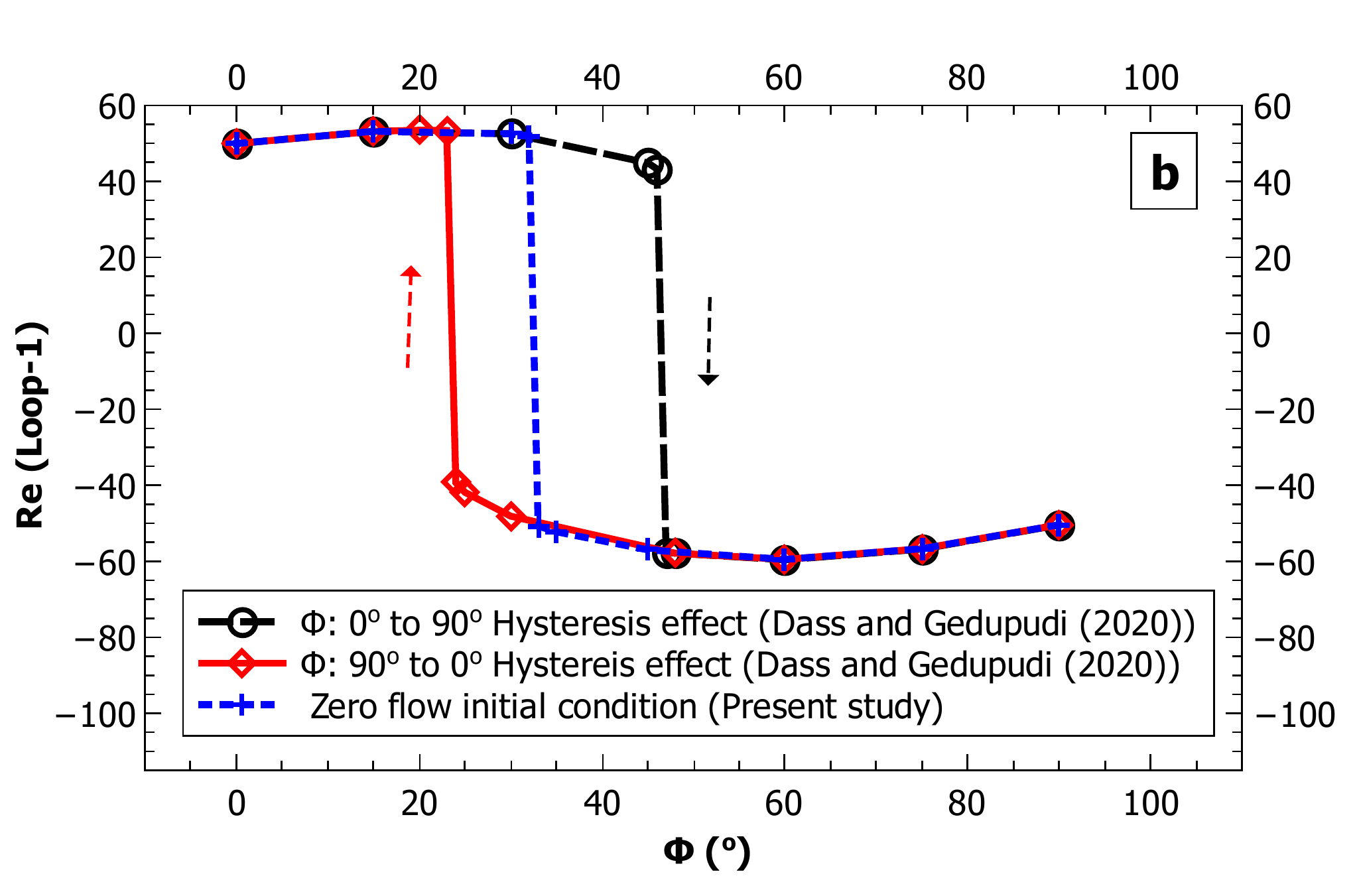}
	\end{subfigure}
	\caption{Effect of $\phi$ on the CNCL. (a) $Nu$ vs $\phi$, (b) $Re (Loop-1)$ vs $\phi$. The notations used to represent the sign of Re is consistent with that used by Dass and Gedupudi \cite{dass2020} for the purpose of comparison.}
	\label{2DCNCLStudy}
\end{figure}

\begin{figure}[!htb]
    \centering
    \includegraphics[width=0.6\linewidth]{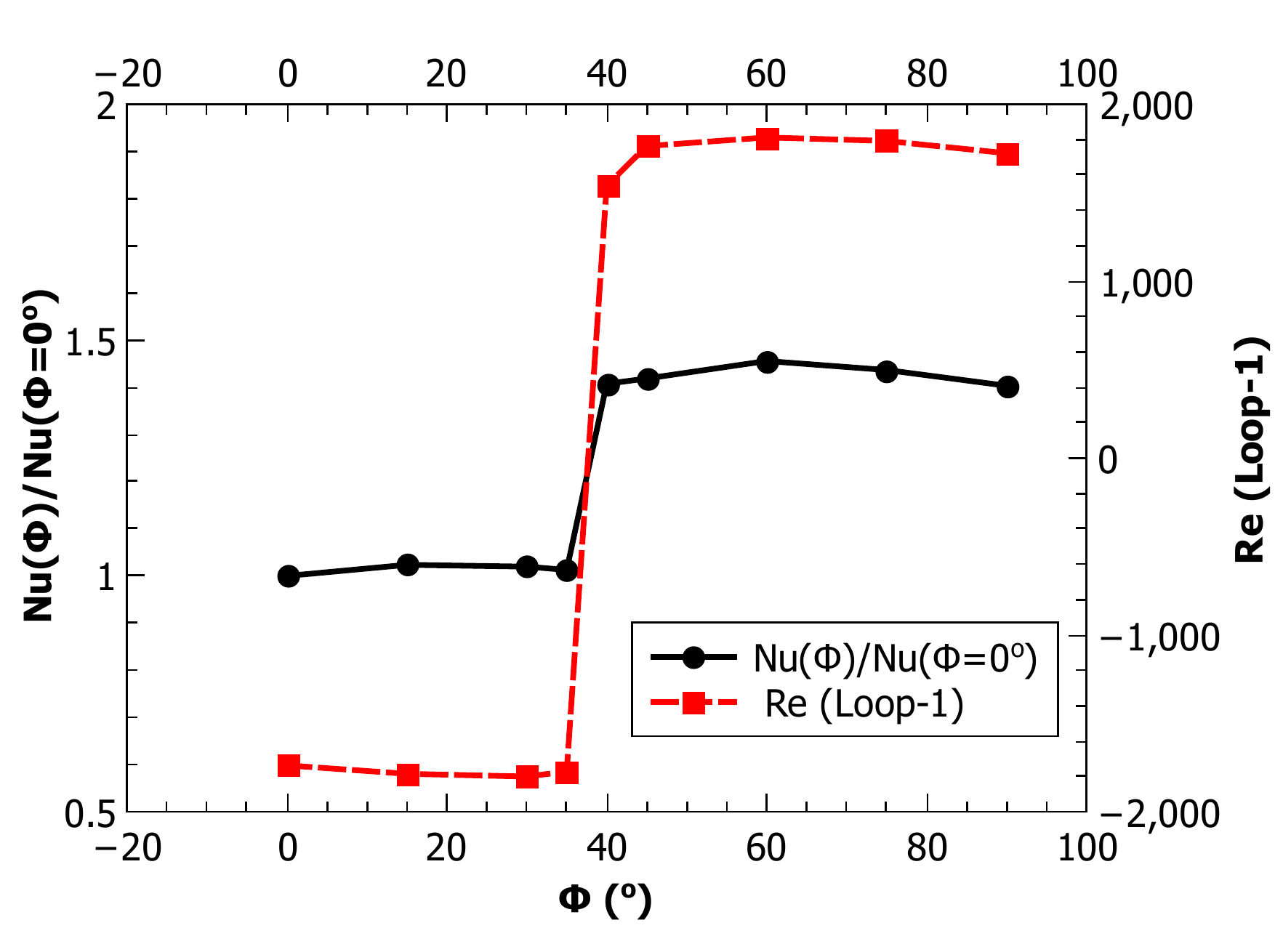}
    \caption{Effect of inclination on the Conjugate CNCL system represented by CCNCL-(d) from 3-D CFD study.}
    \label{3DCCNCLCFDResults}
\end{figure}

To confirm the occurrence of heat transfer jump for the zero flow initial condition the authors conducted a 3-D CFD inclination study for the case of CNCL-(d), the results of which are represented in Figure 12. The jump in heat transfer corresponding to flow direction reversal is clearly witnessed. Thus, it is clear that for the considered Conjugate CNCL case of CCNCL-(d) the heat transfer is a function of inclination. From Figure 12 we can approximate the variation of the Nusselt number ($Nu$) with inclination as a piece-wise function which is expressed as follows:

\begin{equation}
    Nu(\phi)=\left\{
        \begin{array}{ll}
            1 \times Nu (\phi=0^{\circ}), & \quad 0^{\circ}< \phi < 35^{\circ} \\
            1.9 \times Nu (\phi=0^{\circ}), & \quad 35^{\circ} < \phi < 90^{\circ} 
        \end{array}
    \right.
\end{equation}

$Nu(\phi=0^{\circ})$ is calculated using equation (53). 

Incorporating the effect of inclination on the heat transfer coefficient, represented by equation (55), in the 1-D conjugate CNCL model for the case of CCNCL-(d) and comparing the result with 3-D CFD predictions, we obtain Figure 13, which shows a good match between the 1-D model and 3-D CFD data.

\begin{figure}[!htb]
    \centering
    \includegraphics[width=0.6\linewidth]{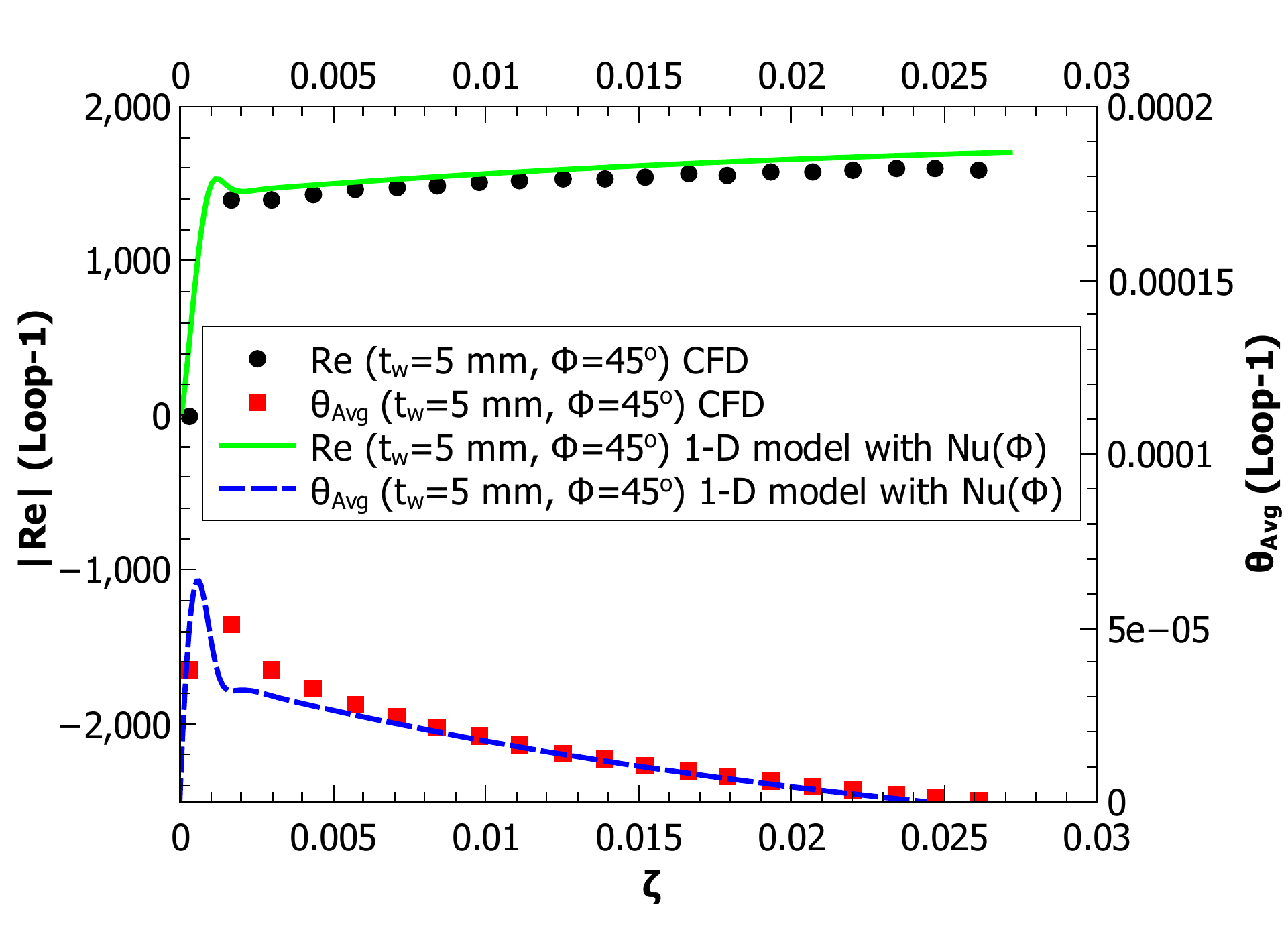}
    \caption{Verification of the case CCNCL-(d) employing the $Nu(\phi)$ correlation.}
    \label{3DCCNCL-(d) verification}
\end{figure}

 Figure 14 shows the comparison between the points of flow direction reversal predicted by the 1-D inclined Conjugate CNCL model and the 3-D CFD. It is observed that there is a slight deviation between the predictions, which may be due to some 3-D effects that have not been incorporated in the 1-D model.

\begin{figure}[!htb]
    \centering
    \includegraphics[width=0.6\linewidth]{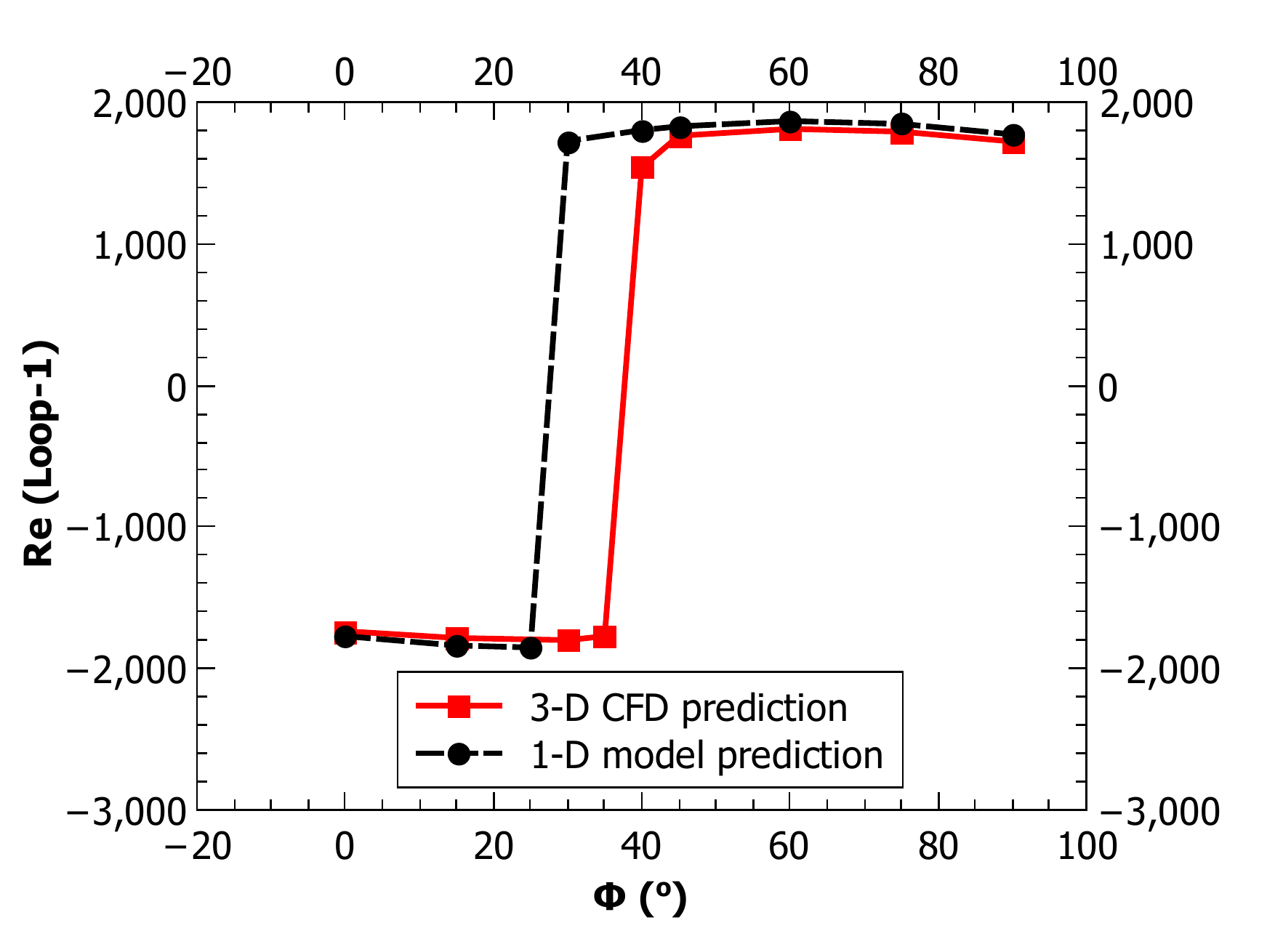}
    \caption{Comparison of the points of flow direction reversal predicted by 3-D CFD and 1-D conjugate CNCL model, for  the case of CCNCL-(d).}
    \label{3DCFDvs1DResults}
\end{figure}

\subsection{Selection of $Co_B$ for modeling $t_w=0$ $mm$ case.}

To model the case CCNCL-(a) which denotes the zero wall thickness case, it is necessary to determine the magnitude of $t_w$ which closely represents the zero wall thickness. This is because as $t_w$ tends to zero, the magnitude of $Co_B$ tends to infinity and we require a finite number as an input to the 1-D Conjugate CNCL model. Figure 15 shows that a magnitude of $t_w=10^{-5}$ $m$ is adequate to capture the transient dynamics of the variables $Re$ and $\theta_{Avg}$. A magnitude of $t_w=10^{-5}$ $m$ corresponds to the magnitude of $Co_B=0.2923$. It is this value which is listed in Table 5 to represent the case corresponding to CCNCL-(a).

\begin{figure}[!htb]
	\centering
	\begin{subfigure}[b]{0.49\textwidth}
		\includegraphics[width=1\linewidth]{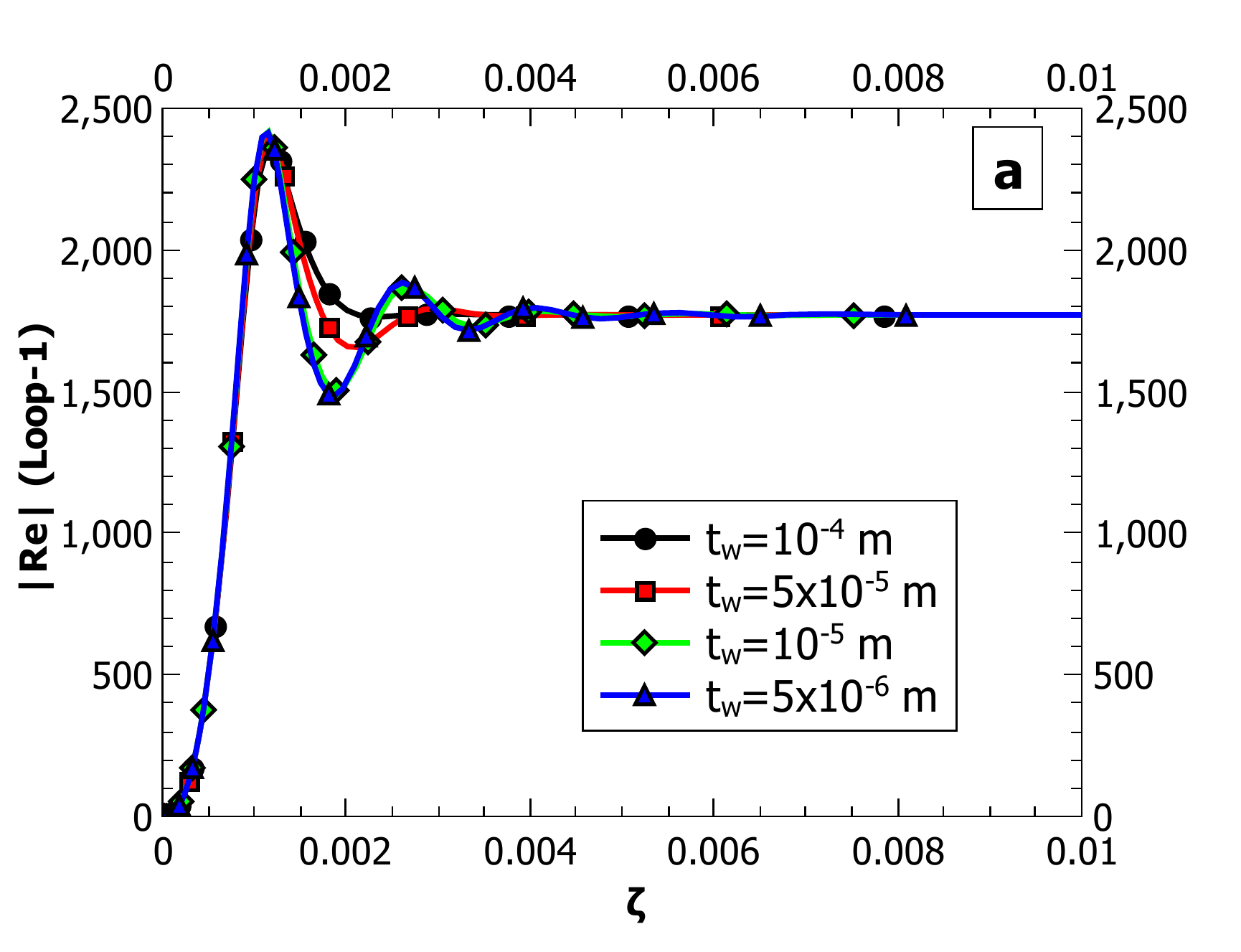}
	\end{subfigure}
	\hspace{\fill}
	\begin{subfigure}[b]{0.49\textwidth}
		\includegraphics[width=1\linewidth]{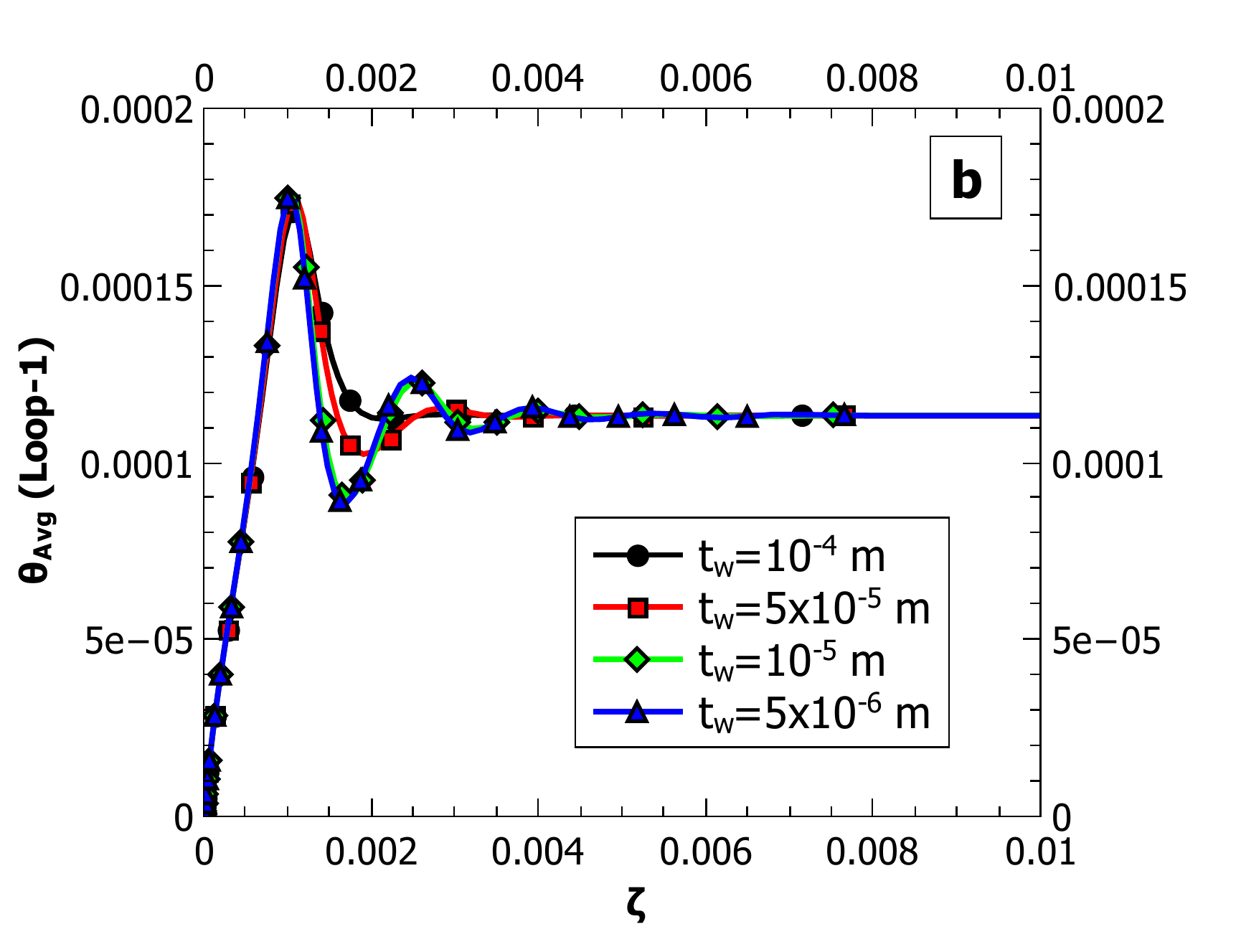}
	\end{subfigure}
	\caption{Selection of appropriate $t_w$ for modelling CCNCL-(a) case for (a) $|Re|$ of Loop 1, (b) $\theta_{Avg}$ of Loop 1.}
	\label{Modeling0mmCase}
\end{figure}

Now that the capability of the 1-D model of the conjugate CNCL to capture the physics has been established, a parametric study is conducted to examine the behaviour of the inclined conjugate CNCL system employing the 1-D model.

\section{Results and Discussion}

A detailed parametric study of the conjugate CNCL system employing the developed 1-D semi-analytical model is presented in this section. Since in the present study, the fluids in the constituent loops of the conjugate CNCL system are identical and for the given heater-cooler position always result in symmetric transient behaviour, it is adequate to represent the dynamics of a single loop of the system to characterize the entire system. The heater cooler position of the considered conjugate CNCL system results in an anticlockwise flow in Loop 1 and a clockwise flow in Loop 2 of the system w.r.t the origin represented in the modelling approach section. The sign associated with the Reynolds number indicates the direction of the flow. The ‘+Ve’ sign indicates clockwise flow direction and ‘-Ve ’ sign anticlockwise flow direction.

A thorough parametric study of a simple CNCL system has been carried out by Dass and Gedupudi \cite{dass2019} for the non-dimensional numbers such as: $Gr$, $Fo$, $Co_A$, $St$ etc. The observations made in the parametric study remain consistent in comparison with the developed 1-D model incorporating the conjugate heat transfer effects, with only a change in the magnitudes. Thus, in the present work the parametric study is confined to the additional new non-dimensional numbers introduced by conjugate effects and to the effects of inclination on the dynamics of conjugate CNCL system.

\subsection{Effect of $Fo_w$}

It is observed from Figure 16a that there is no effect of $Fo_w$ on the initial rise of $Re$ and $\theta_{Avg}$ which is solely governed by the buoyancy forces generated in the heating section for Loop-1 (cooling section for Loop-2). The wall effects begin to assert their influence after the initial rise period and as the system approaches the steady state. From Figure 16a we observe that with increase in the magnitude of $Fo_w$, the rate of heat transfer increases along and across the heat exchanger wall due to increase in thermal conductivity or the non-dimensional number  $Fo_w$. As a result, the system takes lesser time (by around $50\%$, as $Fo_w$ changes from 1 to 15 )  to attain the steady state. Figure 16b represents the steady state trends of $Re$ and $\theta_{Avg}$ with change in $Fo_w$. Thus, it may be inferred that $Fo_w$ contributes to the transient characterization of the system and contributes minimally to the steady state trend of $Re$ and $\theta_{Avg}$. A numerical justification for the lack of a significant relation between $Fo_w$ and parameters $Re$ and $\theta_{Avg}$ at steady state is presented in the second sub-section of the Appendix of the paper. 

\begin{figure}[!htb]
	\centering
	\begin{subfigure}[b]{0.49\textwidth}
		\includegraphics[width=1\linewidth]{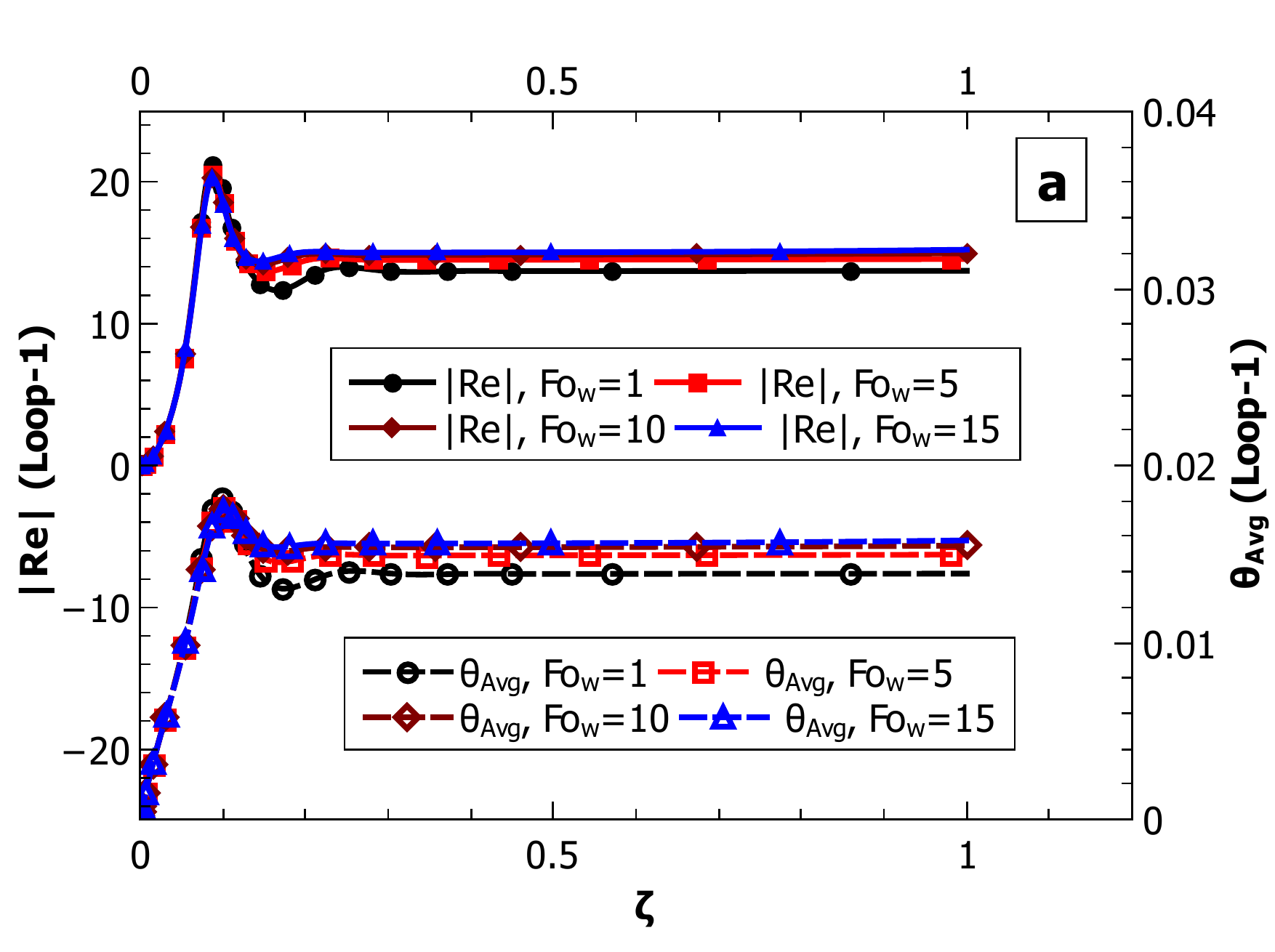}
	\end{subfigure}
	\hspace{\fill}
	\begin{subfigure}[b]{0.49\textwidth}
		\includegraphics[width=1\linewidth]{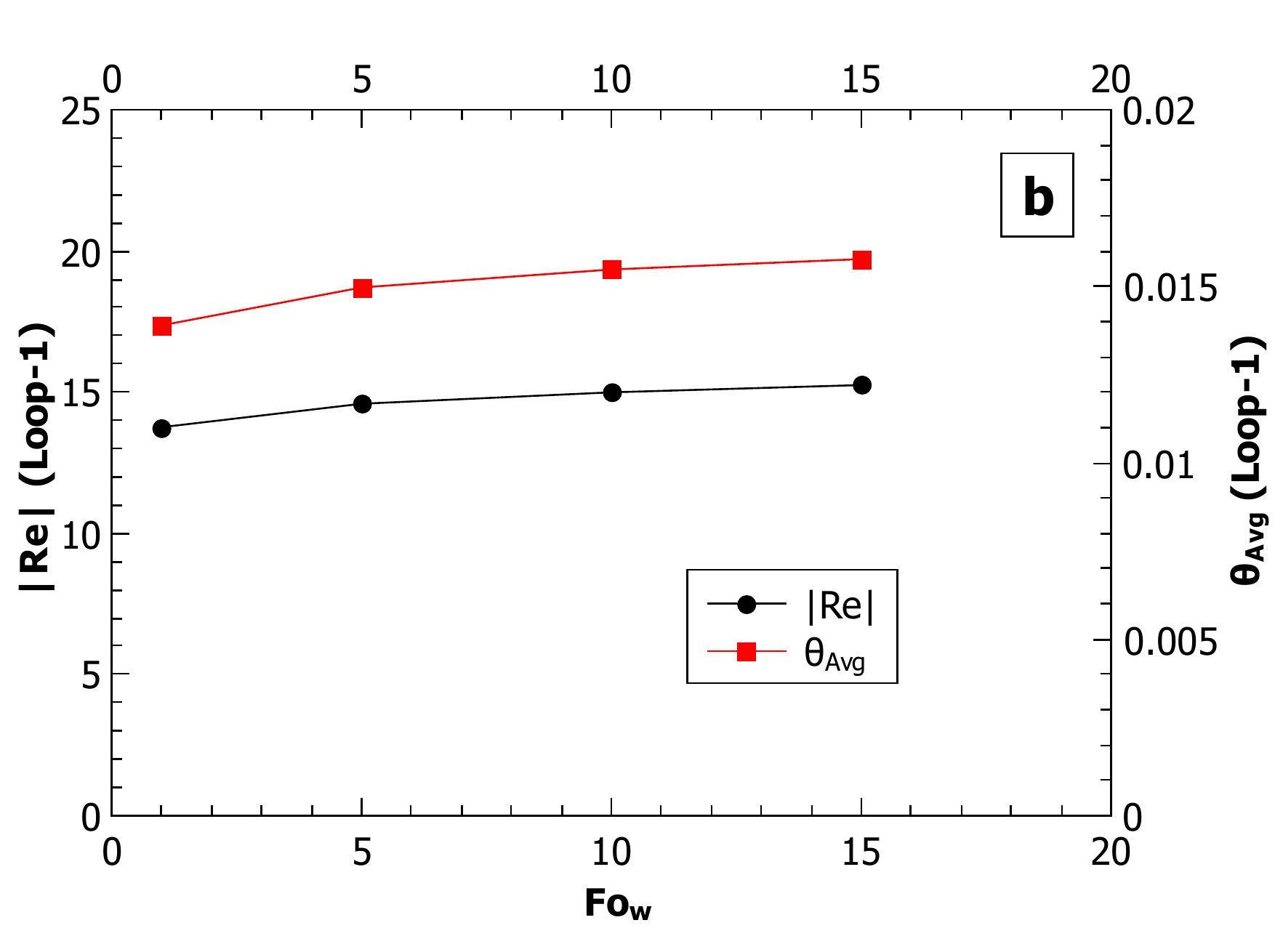}
	\end{subfigure}
	\caption{Effect of Fourier number of the heat exchanger wall ($Fo_w$) on the Conjugate CNCL system for $Gr=10^6$, $Fo=1$, $St=1000$, $Co_A=10$, $Co_B=0.5$, $\phi =0$. (a) transient behaviour, (b) steady state behaviour. }
	\label{EffectOfFow}
\end{figure}

\subsection{Effect of $Co_B$}
$Co_B$ is a non-dimensional parameter which is introduced into the modelling approach to account for the conjugate effects introduced by the heat-exchanger wall of thickness $t_w$. $Co_B$ is inversely proportional to $t_w$ by definition; thus, a lower magnitude of $Co_B$ implies a thicker heat exchanger wall.

\begin{figure}[!htb]
	\centering
	\begin{subfigure}[b]{0.49\textwidth}
		\includegraphics[width=1\linewidth]{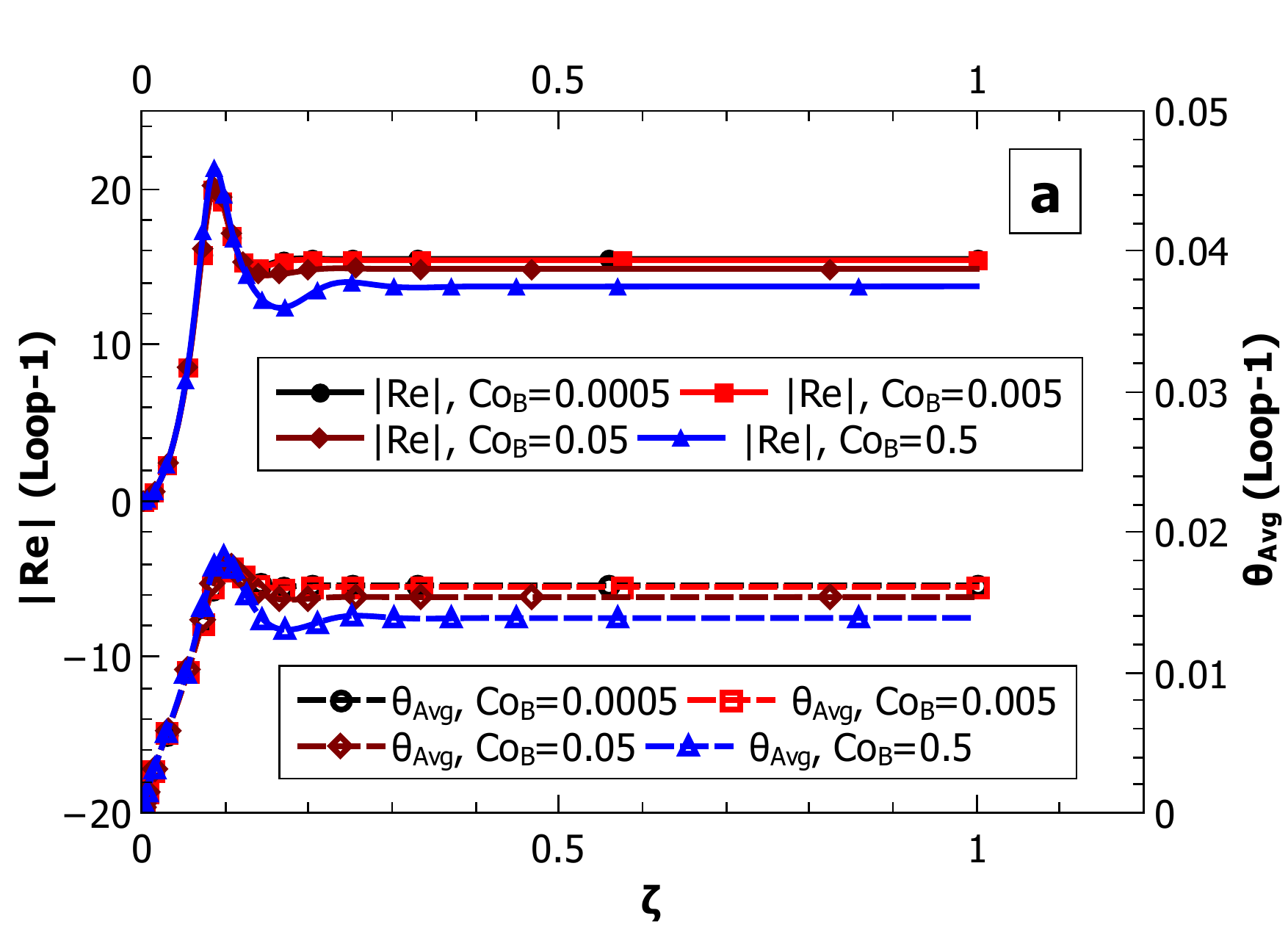}
	\end{subfigure}
	\hspace{\fill}
	\begin{subfigure}[b]{0.49\textwidth}
		\includegraphics[width=1\linewidth]{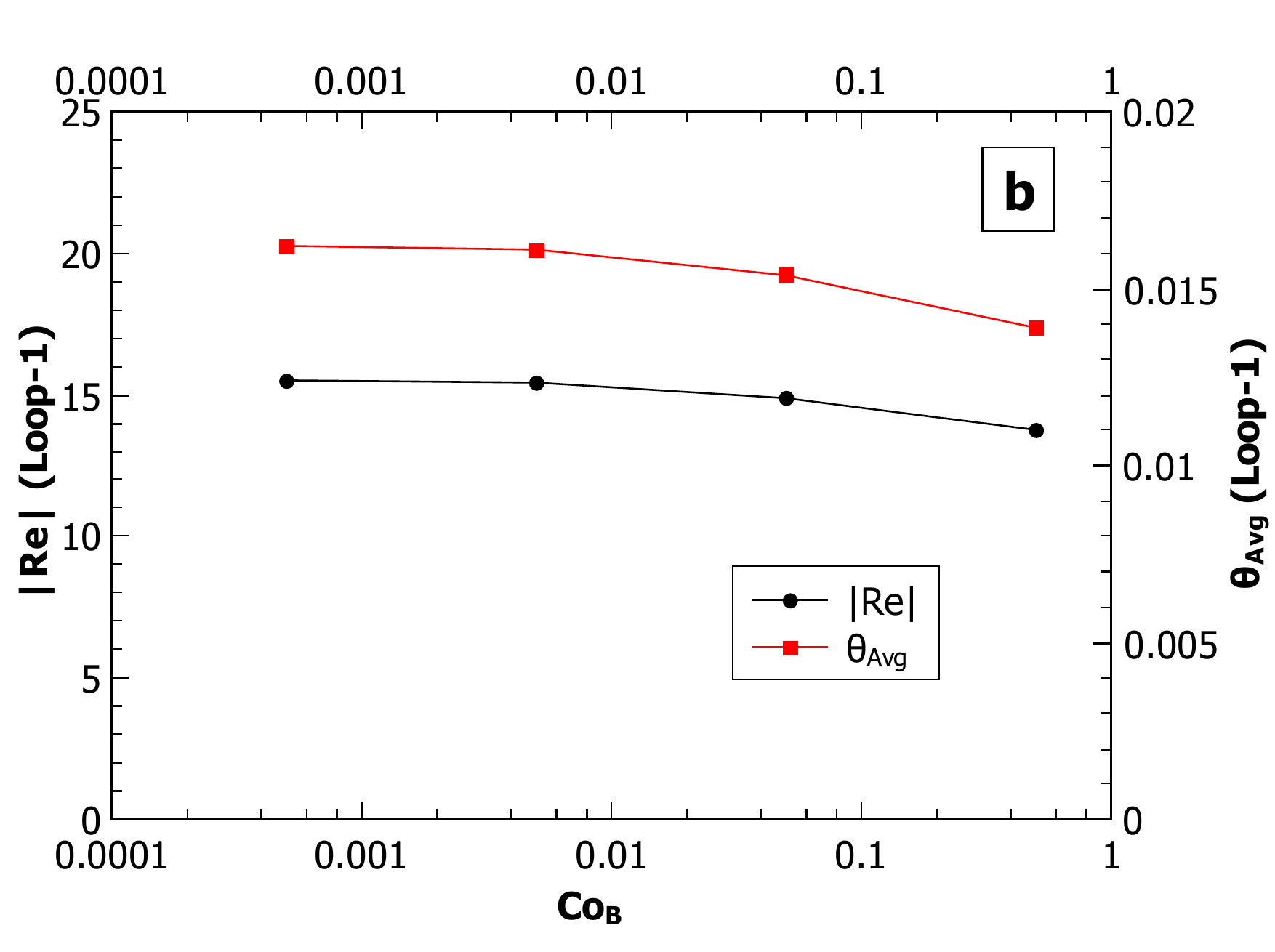}
	\end{subfigure}
	\caption{Effect of $Co_B$ on the Conjugate CNCL system for $Gr=10^6$, $Fo=1$, $St=1000$, $Co_A=10$, $Fo_w=1$, $\phi =0$. (a) transient behaviour, (b) steady state behaviour. }
	\label{EffectOfFow}
\end{figure}

 From Figure 17a, it is observed that with increase in the magnitude of $Co_B$ (decrease in the magnitude of $t_w$) the steady state is attained faster. A small magnitude of $Co_B$ corresponds to a larger wall thickness, which also leads to a larger thermal inertia of the wall that tends to suppress the oscillatory transience of the conjugate CNCL system. From Figure 17b it is noted that variation in $Co_B$ does not have a significant influence on the steady state magnitudes of $Re$ and $\theta_{Avg}$. A numerical justification for the lack of a significant relation between $Co_B$ and parameters $Re$ and $\theta_{Avg}$ at steady state is presented in the second sub-section of the Appendix of the paper. 

\subsection{Effect of $St$}

Figure 18a and Figure 18b show the effect of Stanton number on the conjugate CNCL system. Figure 16a indicates that a change in the magnitude of $St$ does not affect the transient nature of the Reynolds number of the system, but it is observed that $\theta_{Avg}$ is sensitive to the change in $St$. The increase in magnitude of $St$ results in a drop in the magnitude of $\theta_{Avg}$. This trend is different from that reported by Dass and Gedupudi (2019) \cite{dass2019} who observed very low sensitivity of both $Re$ and $\theta_{Avg}$ to change in $St$. This difference is due to the fact that the range of $St$ considered for the present study is between $100-1000$ and the range of $St$ considered by Dass and Gedupudi \cite{dass2019} is between $1000-5000$ and from figure 18b it is clear that the dependence of $\theta_{Avg}$ decreases with the increase in $St$. A numerical justification for the lack of a significant relation between $St$ and parameter $Re$ is presented in the second sub-section of the Appendix of the paper. 

\begin{figure}[!htb]
	\centering
	\begin{subfigure}[b]{0.49\textwidth}
		\includegraphics[width=1\linewidth]{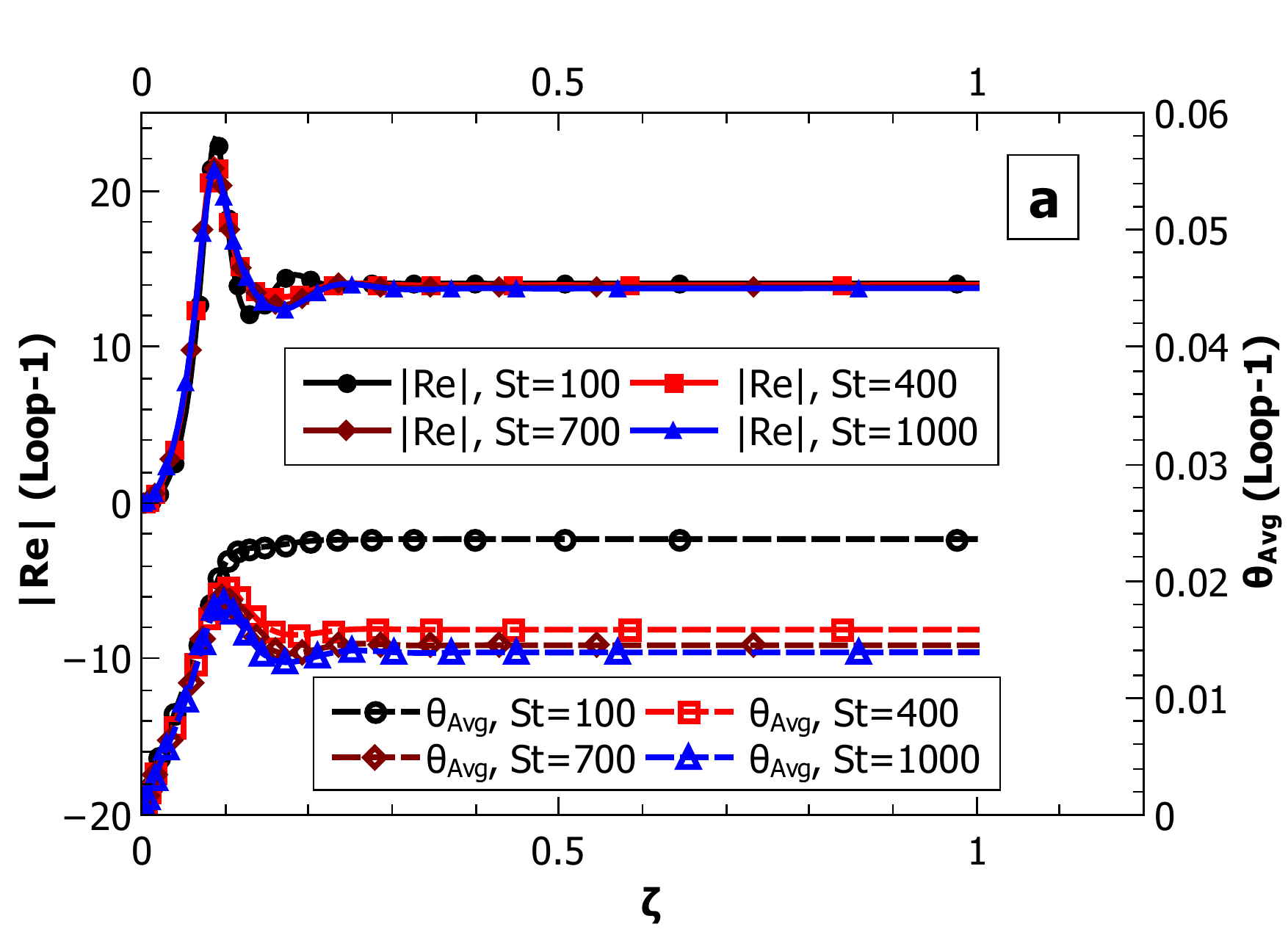}
	\end{subfigure}
	\hspace{\fill}
	\begin{subfigure}[b]{0.49\textwidth}
		\includegraphics[width=1\linewidth]{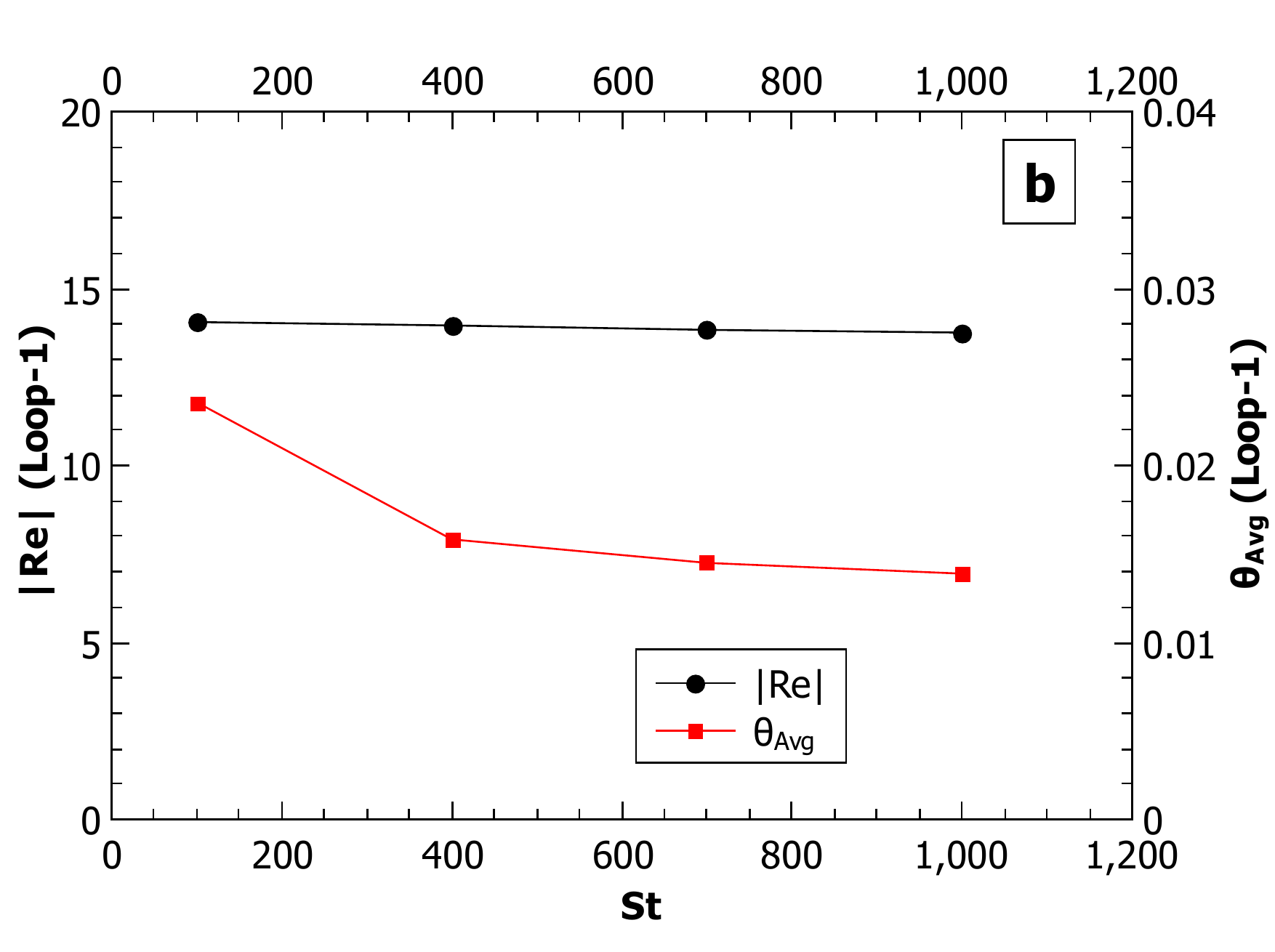}
	\end{subfigure}
	\caption{Effect of $St$ on the Conjugate CNCL system for $Gr=10^6$, $Fo=1$, $Co_A=10$, $Fo_w=1$, $\phi =0$. (a) transient behaviour for $Co_B=0.5$,  (b) steady state behaviour for $Co_B=0.5$, $Co_B=0.0005$. }
	\label{EffectOfSt}
\end{figure}

\subsection{ Effect of inclination ($\phi$) on the Conjugate CNCL system}

\begin{figure}[!htb]
	\centering
	\begin{subfigure}[b]{0.49\textwidth}
		\includegraphics[width=1\linewidth]{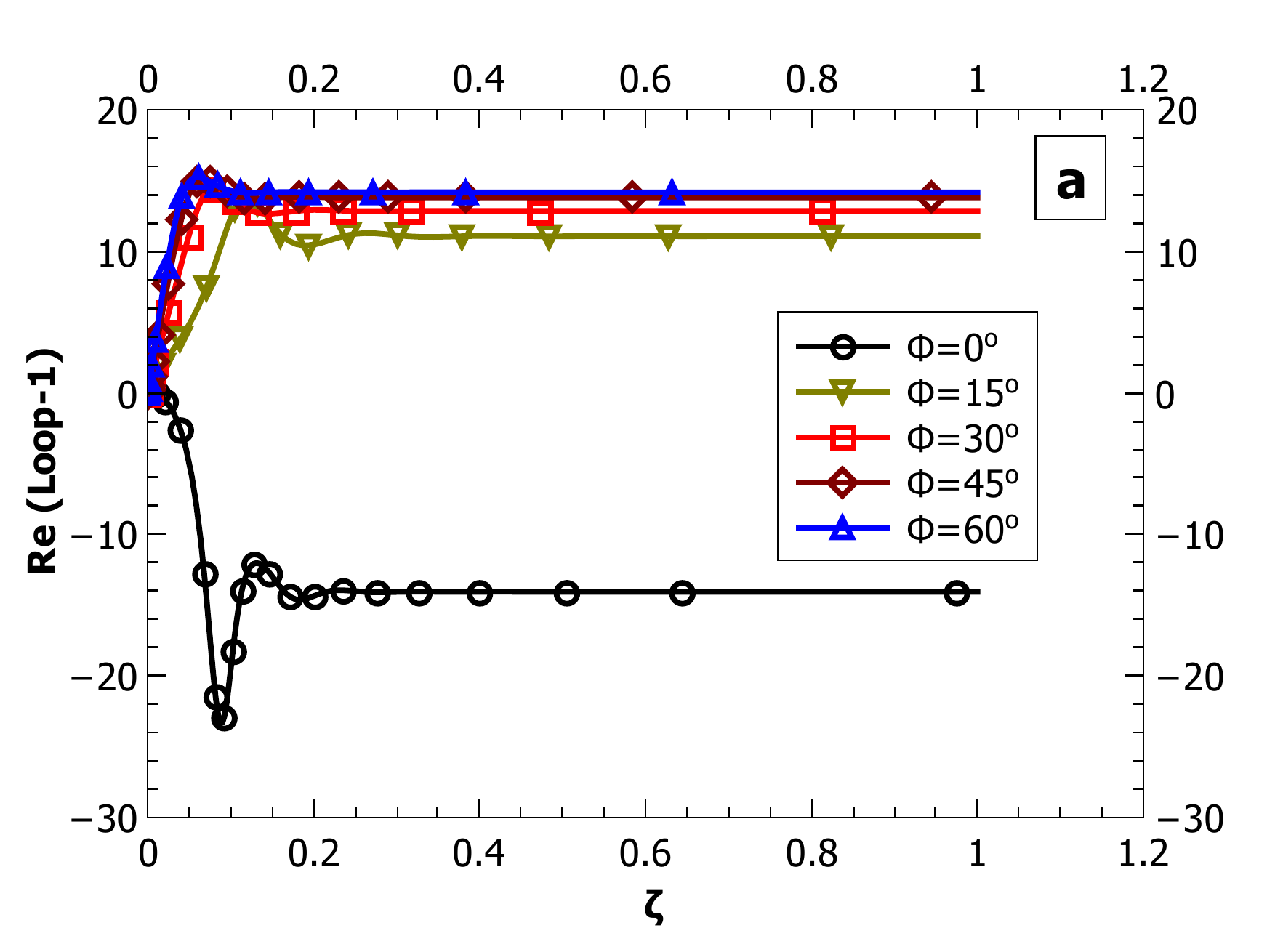}
	\end{subfigure}
	\hspace{\fill}
	\begin{subfigure}[b]{0.49\textwidth}
		\includegraphics[width=1\linewidth]{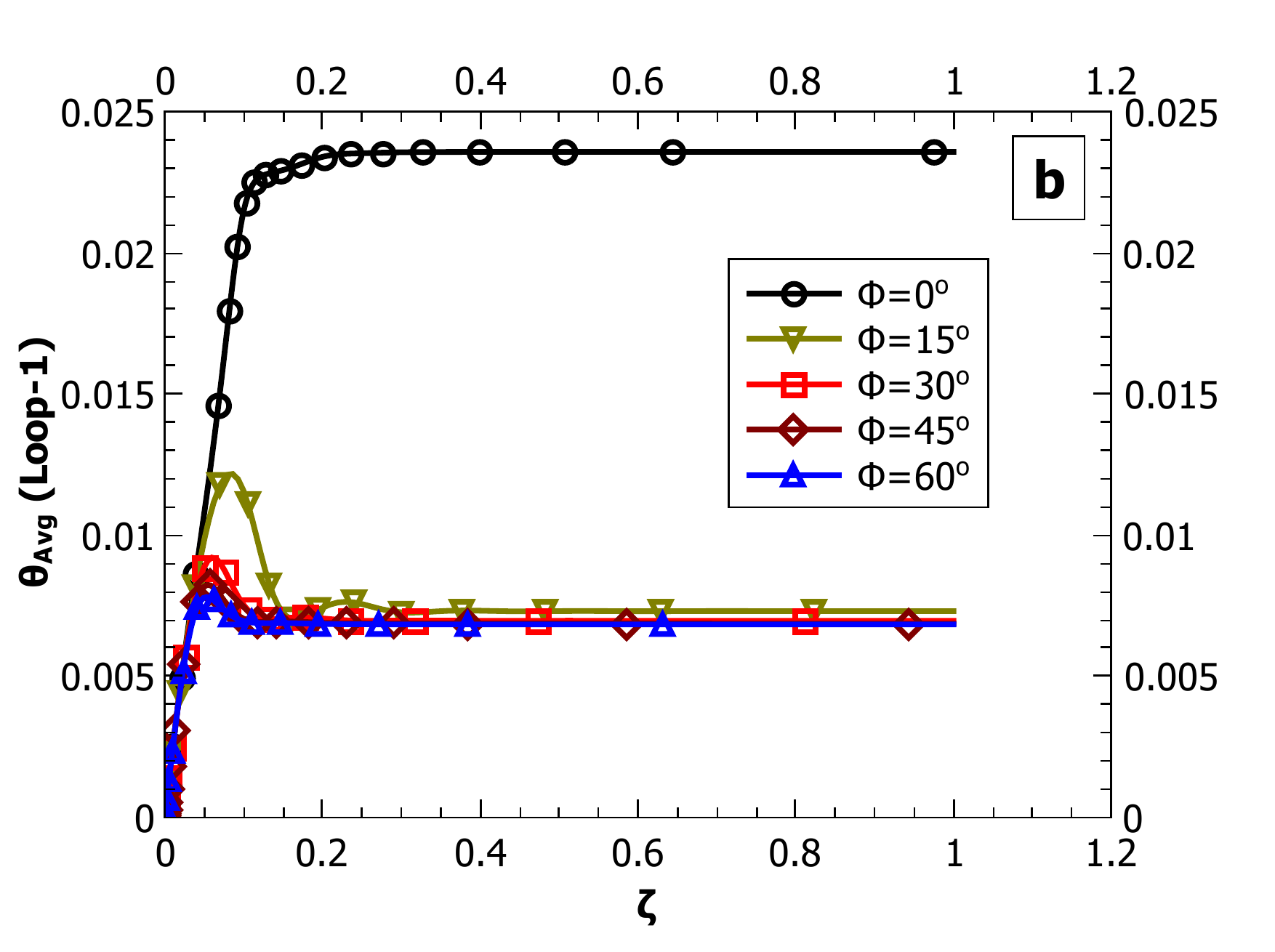}
	\end{subfigure}
	\caption{Effect of $\phi$ on the transient behaviour of Loop 1 of the Conjugate CNCL system for $Gr=10^6$, $Fo=1$, $Co_A=10$, $Co_B=0.5$, $Fo_w=1$, $St =100$. (a) Reynolds number ($Re$), (b) Non-dimensional average fluid temperature ($\theta_{Avg}$). }
	\label{EffectOfPhi}
\end{figure}

The effect of inclination on the conjugate CNCL system is presented in Figure 19a and Figure 19b. From Figure 19a, which represents the effect of inclination on the transient $Re$ behaviour, we observe that the flow direction flips from anti-clockwise to clockwise with introduction of inclination to the conjugate CNCL system. This happens due to the shift in the location where buoyancy forces have the largest magnitude. At $\phi=0$, the cooling section of the system determines the flow direction as the heating section is horizontal w.r.t gravity, but with the introduction of inclination the buoyancy forces generated at the heating section force the fluid in the clockwise direction leading to the observed behaviour. Figure 19b shows a huge deviation between the transient behaviour of $\theta_{Avg}$ at $\phi=0$ and that at other angles ($15^{\circ}$, $30^{\circ}$, $45^{\circ}$, $60^{\circ}$), which is also due to the reversal in the flow direction that leads to the change in the temperature profile along the length of the component NCL of the inclined conjugate CNCL system.

\subsection{Effect of $Fo_w$ and $Co_B$ on the flow direction reversal}

Figure 20 represents the effects of $Fo_w$ and $Co_B$ on the point of flow direction reversal induced by the change in the inclination of the system at steady state. It is observed that both the parameters do not influence the point at which the flow direction reversal occurs. This observation is in accordance of the fact that $Fo_w$ and $Co_B$ do not alter the steady state magnitude of $Re$ as reported in the earlier sections. The point at which the flow direction reversal occurs can be determined by the stability analysis which is beyond the scope of the present study.  

\begin{figure}[!htb]
	\centering
	\begin{subfigure}[b]{0.49\textwidth}
		\includegraphics[width=1\linewidth]{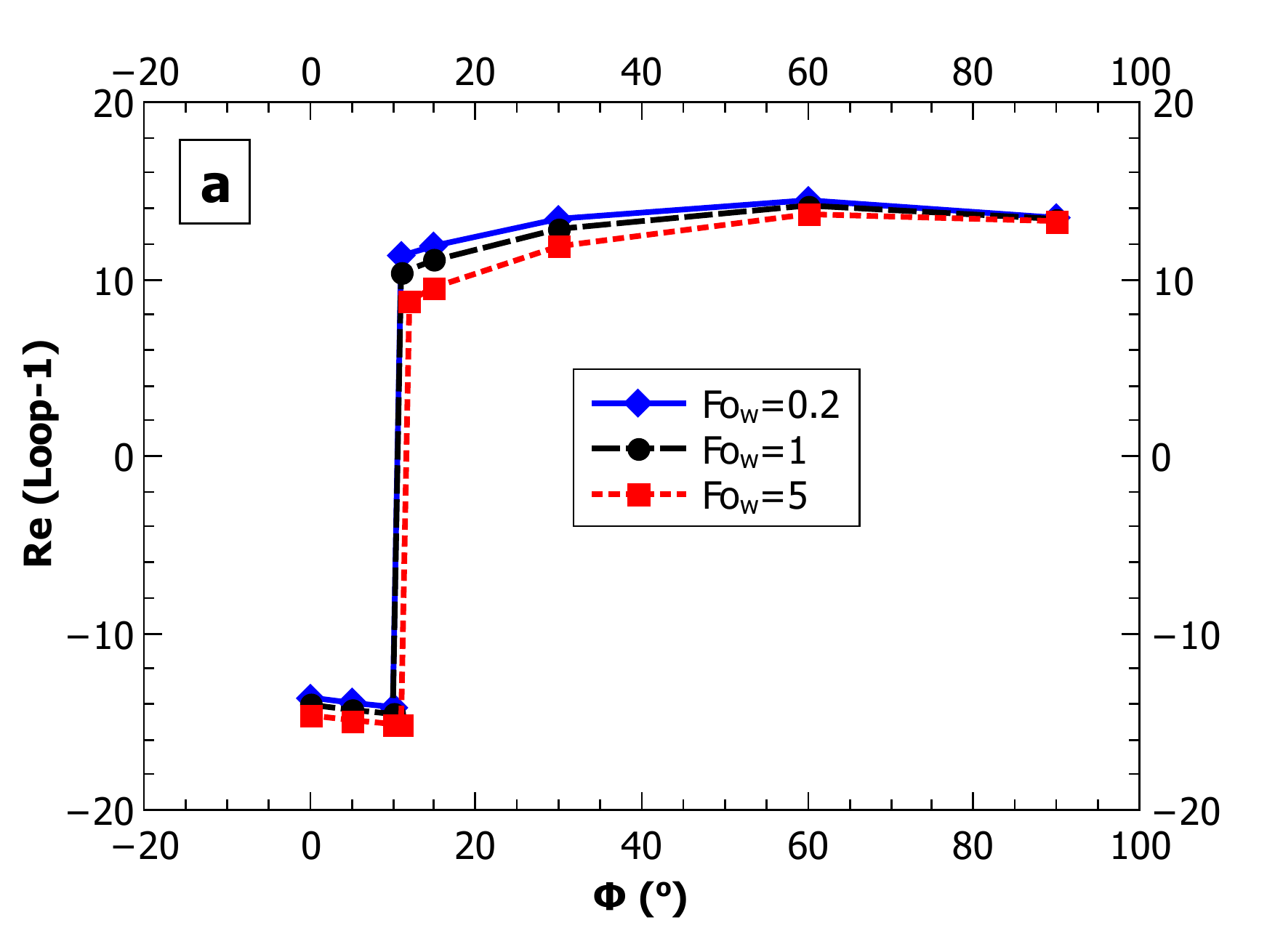}
	\end{subfigure}
	\hspace{\fill}
	\begin{subfigure}[b]{0.49\textwidth}
		\includegraphics[width=1\linewidth]{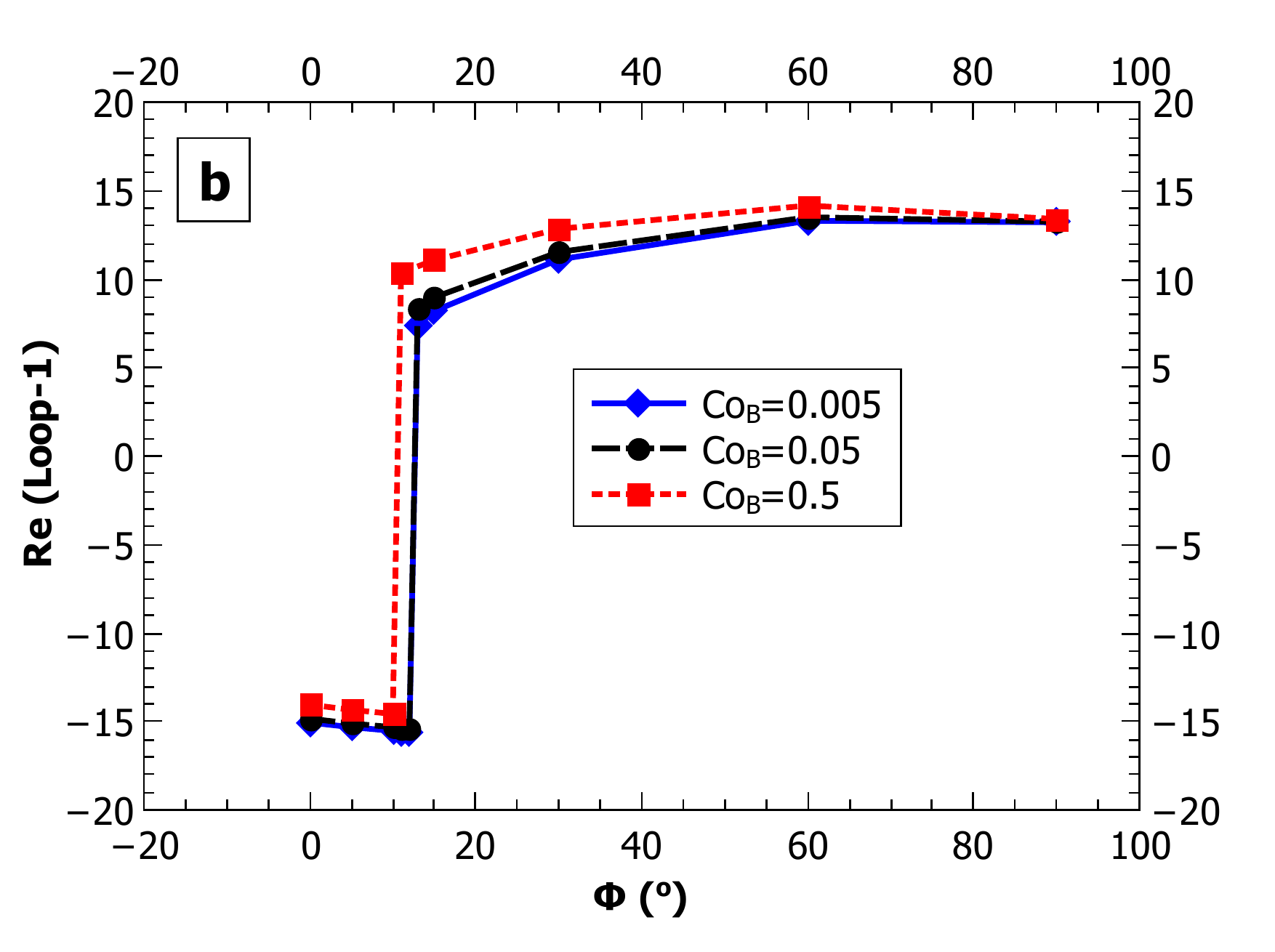}
	\end{subfigure}
	\caption{Effect on the flow direction reversal in inclined conjugate CNCL system for $Gr=10^6$, $Fo=1$, $Co_A=10$, $St =100$ by parameter (a) $Fo_w$ with $Co_B=0.5$, (b) $Co_B$ with $Fo_w=1$. }
	\label{Effect of Fow and CoB on flow reversal}
\end{figure}

\section{Steady state analysis of the Conjugate CNCL system}

To thoroughly understand the parametric study presented in the preceding sections quantitatively, a steady state analysis of the Conjugate CNCL system is carried out. To simplify the analysis, the temperature variation in the diabatic sections of the system are assumed to be linear. Figure 21 depicts the steady state non-dimensional temperature profile of Loop 1, Wall and Loop 2 incorporating the linear profile assumption. $m_1$ and $m_2$ are the slopes of the non-dimensional temperature profiles in common heat exchanger section of the conjugate CNCL system. 

\begin{figure}[!htb]
    \centering
    \includegraphics[width=0.8\linewidth]{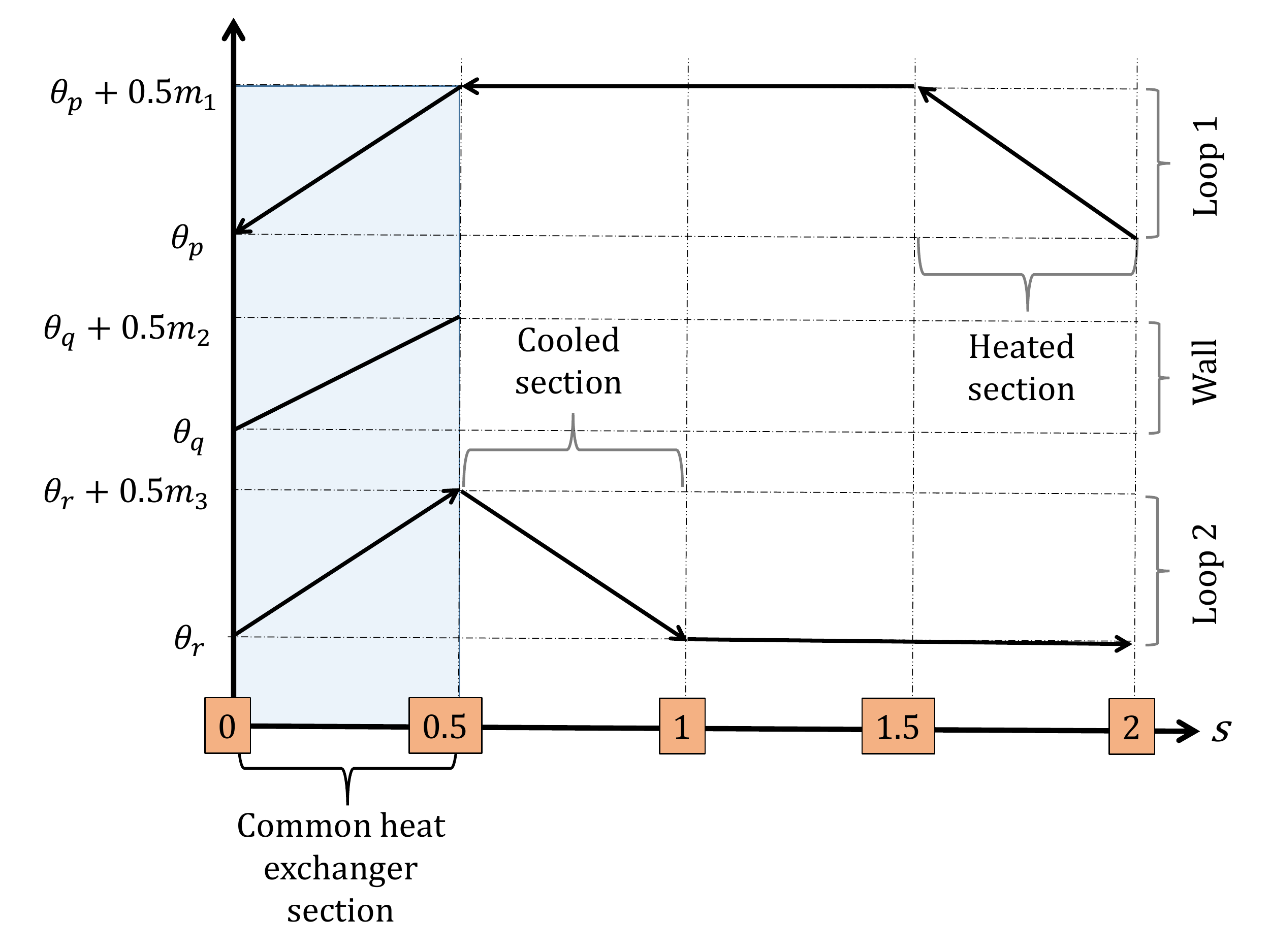}
    \caption{Steady state non-dimensional temperature profile of the Conjugate CNCL system. The arrows indicate the direction of fluid flow in Loop 1 and Loop 2 of the Conjugate CNCL system.}
    \label{SteadyStateAnalysis}
\end{figure}

The non-dimensional temperature profiles represented in  Figure 21 can be represented mathematically as follows:

\begin{equation}
    \theta_1(s)=\left\{
        \begin{array}{llll}
            \theta_p + m_1s, & \quad 0< s < 0.5 \\
            \theta_p + 0.5m_1, & \quad 0.5< s < 1 \\
            \theta_p + 0.5m_1, & \quad 1< s < 1.5 \\
             \theta_p + 2m_1 - m_1s, & \quad 1< s < 1.5
        \end{array}
    \right.
\end{equation}

\begin{equation}
    \theta_w(s)=\left\{
        \begin{array}{ll}
            \theta_q + m_2s, & \quad 0< s < 0.5 \\
            0, & \quad 0.5< s < 2 
        \end{array}
    \right.
\end{equation}

This implies that $\frac{d^2 \theta_1}{ds^2}=\frac{d^2 \theta_1}{ds^2}=0$. This is a consequence of the linear profile assumption made earlier.

\subsection{Parameters influencing $Re_1$ at steady state}

The steady state momentum equation of Loop 1 of the Conjugate CNCL system is:

\begin{equation}
 Co_A\; Re_1^{(2-d)} = Gr\oint \theta_1 \;f(s)ds - \frac{NK}{4}Re_1^2    
\end{equation}

Evaluating the non-dimensional temperature integral from Figure 18 for $\phi=0^{\circ}$, we get:

\begin{equation}
    \oint \theta_1 \;f(s)ds= -\frac{m_1}{8} 
\end{equation}

Using equations (58) and (59), it can be concluded that $Re$ is a function of $m_1$, $Gr$ and $Co_A$ ($Re=f(m_1,Gr,Co_A)$).

The steady state energy equation of Loop 1 is as follows:

\begin{equation}
    Re_1\frac{d \theta_1}{d s} = h_1(s) -\lambda(s) St_1 (\theta_1 - Co_B\; \theta_w) + Fo_1\frac{d^2 \theta_1}{d s^2}
\end{equation}

The energy provided at the heating section at steady state is equivalent in magnitude to the energy transferred by convection, which yields the following equation:

\begin{equation}
    Re_1\frac{d \theta_1}{d s} = h_1(s)
\end{equation}

Evaluating equation (61) by substituting the non-dimensional temperature at the heated section yields the following equation:

\begin{equation}
    Re_1m_1 = 1
\end{equation}

From earlier inference we know that $Re=f(m_1,Gr,Co_A)$, which together with equation (62) results in the inference that as long as parameters $Gr_1$,$Co_A$ are kept constant (as is done for the parametric studies in the previous section) the magnitude of $Re$ is unaltered. This observation is in accordance with the conclusions from the parametric study that $Re$ is not a function of $Fo_w$, $Co_B$ and $St$.

\subsection{Parameters influencing $\theta_{1,Avg}$ at steady state}

The magnitude of energy supplied at the heating section is transferred to the common heat exchange section and is equivalent in magnitude, which yields the following relation:

\begin{equation}
   \oint \lambda(s) St_1 (\theta_1 - Co_B\; \theta_w) ds =\oint h_1(s) ds
\end{equation}

Evaluating equation (63) by substituting the non-dimensional temperature profile corresponding to the common heat exchange section yields:

\begin{equation}
    St_1[ 4(\theta_p - Co_B\; \theta_q) + (m_1 - Co_B\; m_2)] =4
\end{equation}

The steady state equation of the wall is :

\begin{equation}
    0 =\lambda(s) St_1 (\theta_1 + \theta_2 -2 \;Co_B\; \theta_w) + \lambda_s(s)Fo_w\frac{d^2 \theta_w}{d s^2}
\end{equation}

The energy transmitted by the Loop 1 to the wall is equivalent in magnitude to the energy transferred by the wall to Loop 2. This observation can be expressed mathematically as :

\begin{equation}
    \oint \lambda (s) St_1(\theta_1 - Co_B\; \theta_w) = \oint \lambda (s) St_1( Co_B\; \theta_w - \theta_2)
\end{equation}

\begin{equation}
    \implies \theta_{w,Avg}=\frac{\theta_{1,Avg} + \theta_{2,Avg}}{2\;Co_B}
\end{equation}

From the work done by Dass and Gedupudi \cite{dass2019} for a CNCL system for different heater cooler configurations, the following relation always holds :

\begin{equation}
    \frac{\theta_{1,Avg}}{\theta_{2,Avg}}=-1
\end{equation}

Equation (68) is a consequence of symmetry of the CNCL system, as the Conjugate CNCL is also symmetric, equation (68) also holds true for it. Substituting equation (68) in equation (67) yields:

\begin{equation}
     \theta_{w,Avg}=0
\end{equation}

It is noted that:

\begin{equation}
     \theta_{1,Avg}=\oint \theta_1(s) ds = \theta_p + 0.25m_1
\end{equation}

\begin{equation}
     \theta_{w,Avg}=\oint \theta_w(s) ds = \theta_q + 0.25m_2
\end{equation}

Solving equations (64),(69),(70) and (71),  we obtain:

\begin{equation}
    St_1\;\theta_{1,Avg}=1
\end{equation}

Equation (72) implies that $\theta_{1,Avg}$ is independent of $Fo_w$ and $Co_B$ and is inversely proportional to $St$. This is accordance with the observations from the parametric studies conducted in the previous section.

\section{Conclusions}

The current work presents the 1-D modelling of a conjugate CNCL system employing the Fourier series-based approach. The modelling incorporates the effect of wall conduction in the heat exchanger wall and the inclination on the system. The developed 1-D model is found to be in good agreement with the 3-D CFD studies performed to verify the model. The following conclusions can be drawn from the study:
\begin{enumerate}
    \item It can be clearly observed from the study that the inclusion of the wall conduction effects in the study has a significant impact on the transient behaviour of the system and thus must not be neglected.
    \item The inclusion of conduction effects in the modelling introduces two new non-dimensional numbers into the study, namely $Fo_w$ and $Co_B$. The increase in the magnitude of $Fo_w$ and the decrease in the magnitude of $Co_B$ lead to the quicker attainment of steady state of the conjugate CNCL system. $Fo_w$ and $Co_B$ do not influence the steady state magnitude of $Re$ and $\theta_{Avg}$ significantly.
    
    \item The Stanton number ($St$) of the conjugate CNCL system is identified as an important parameter which influences the transient dynamics of the non-dimensional temperature of the conjugate CNCL system. The non-dimensional temperature is inversely proportional to $St$. The variation in the Stanton number has negligible influence on the transient and steady state behaviour of $Re$.
    \item The inclination of the system is observed to be a parameter that influences the flow direction. From the present 3-D CFD study, it is observed that the considered system displays a jump in the heat transfer coefficient with variation in inclination. This jump in the heat transfer coefficient with inclination also occurs with zero initial flow conditions (no hysteresis effect). The jump occurs at the point at which flow direction reversal occurs. The determination of suitable heat transfer coefficient correlation considering inclination effects for conjugate CNCL systems will be part of the future study.
    \item The 1-D model of the conjugate CNCL system is capable of predicting the flow direction reversal with varying inclination, but there is a slight deviation between the predictions made by the 1-D model and the 3-D CFD with regard to the angle at which the flow direction reversal occurs. This deviation may be due to the influence of 3-D effects.
    \item The parameters $Fo_w$ and $Co_B$ do not shift the point at which the flow direction reversal occurs.
    \item The 1-D model developed in the present study is restricted to systems with aspect ratio ($L/L1$) unity. This is done to simplify the evaluation of boundary condition at the wall. The model may be generalised for systems of different aspect ratios easily by recalculating the appropriate boundary conditions.
    \item The developed 1-D model is an ideal tool to quickly model the transient dynamics of the inclined conjugate CNCL system and to estimate the time required for the system to attain steady state, compared to the 3-D CFD studies which are computationally expensive. 
\end{enumerate}

\section{Appendix}

\subsection{Derivation of the energy equation of the wall}

The control volume of the wall used for derivation of governing equations is represented in Figure 22.

\begin{figure}[!htb]
    \centering
    \includegraphics[width=0.4\linewidth]{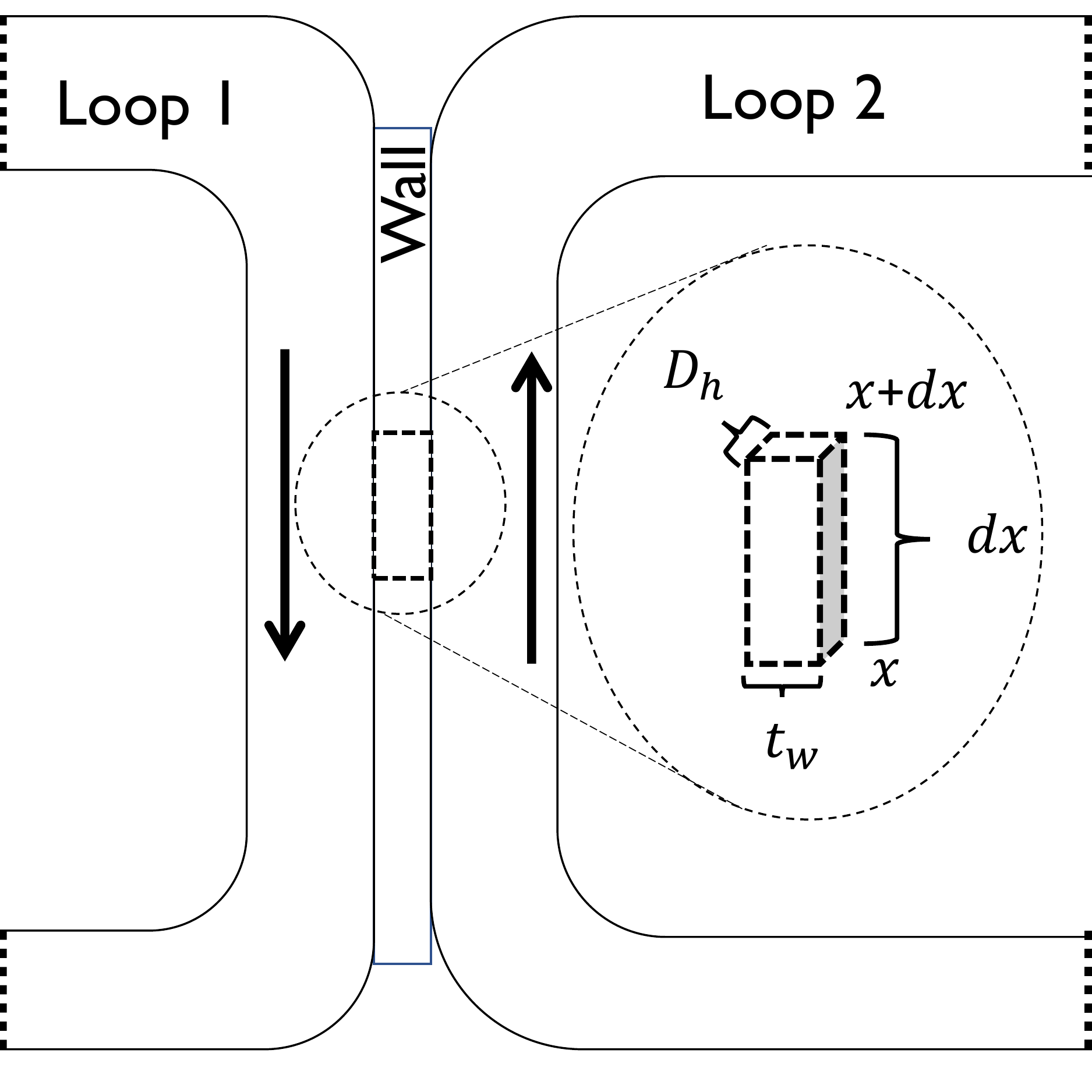}
    \caption{Control volume used for deriving the wall energy equation}
    \label{WallEnergyEquation}
\end{figure}

Area of the cross section of the control volume ($A_{cs}$)= $t_w\; D_h$

Area of surface of control volume in contact with Loop 1 ($A_1$)=$D_h \; dx$

Area of surface of control volume in contact with Loop 2 ($A_2$)=$D_h \; dx$

Net rate of change of energy of the control volume = $\rho_w C_{p,w} A_{cs} dx \frac{dT_w}{dt}$

Energy absorbed by the control volume = $U A_1 \lambda (x) (T_1 -T_w)$

Energy released by the control volume = $U A_2 \lambda (x) (T_w -T_s)$

Heat transfer by conduction within the control volume= $\kappa_w \lambda(x) A_{cs} \frac{\partial^2 T_w}{\partial x^2}$

Applying energy balance to the control volume yields:

\begin{equation}
    \rho_w C_{p,w} A_{cs} dx \frac{dT_w}{dt}=U A_1 \lambda (x) (T_1 -T_w) - U A_2 \lambda (x) (T_w -T_2) + \kappa_w \lambda(x) A_{cs} \frac{\partial^2 T_w}{\partial x^2}
\end{equation}

Simplifying the above equation yields:

\begin{equation}
    \frac{\partial T_w}{\partial t}=\frac{U}{\rho_w C_{p,w}t_w} \lambda (x) (T_1 + T_w -2T_w) + \alpha_w \lambda(x)  \frac{\partial^2 T_w}{\partial x^2}
\end{equation}

Equation 74 represents the energy equation of the wall.

\subsection{Sensitivity of $Re_{ss}$ and $\theta_{Avg,ss}$ to $Fo_w$, $Co_B$ and $St$}

To assess the sensitivity of $Re_{ss}$ and $\theta_{Avg,ss}$ to the changes in $Fo_w$, $Co_B$ and $St$, the following parameters are employed:

$X: Fo_w,\; Co_B,\; St$, $Y: Re_{ss},\;\theta_{Avg,ss}$

$\% \Delta X$= Percentage change in the magnitude of parameter $X$ with the previous considered value.

$\% \Delta Y$= Percentage change in the magnitude of parameter $Y$ with the previous considered value.

$\frac{\% \Delta Y}{\% \Delta X}$ = This ratio determines the sensitivity of para mater $Y$ to parameter $X$, or it represents the percentage change induced in para mater $Y$ by a $1\%$ change in magnitude of para mater $X$.

Table 6 represents the sensitivity of $Re_{ss}$ and $\theta_{Avg,ss}$ to $Fo_w$. The values of $Fo_w$, $Re_{ss}$ and $\theta_{Avg,ss}$ are obtained from Figure 16b. It is observed from Table 6 that a $1\%$ change in $Fo_w$ results in a maximum of $\approx 0.03\%$ change in the magnitudes of $Re_{ss}$ and $\theta_{Avg,ss}$, indicating a negligible dependence of steady state parameters on $Fo_w$.

\begin{table}[!htb]
\centering
\caption{Sensitivity of $Re_{ss}$ and $\theta_{Avg,ss}$ to $Fo_w$}
\begin{tabular}{|l|l|l|l|l|l|l|l|}
\hline
$Fo_w$ & $Re_{ss}$ & $\theta_{Avg,ss}$ & $\% \Delta Fo_w$ & $\% \Delta Re_{ss}$ & $\% \Delta \theta_{Avg,ss}$ & $\frac{\% \Delta Re_{ss}}{\% \Delta Fo_w}$ & $\frac{\% \Delta \theta_{Avg,ss}}{\% \Delta Fo_w}$ \\ \hline
1    & 13.75766  & 0.013891          &                &                     &                             &                                          &                                                  \\ \hline
5    & 14.58791  & 0.014965          & 400            & 6.03481             & 7.727645                    & 0.015087                                 & 0.019319                                         \\ \hline
10   & 14.99249  & 0.015474          & 100            & 2.77342             & 3.406638                    & 0.027734                                 & 0.034066                                         \\ \hline
15   & 15.25041  & 0.015769          & 50             & 1.720355            & 1.904287                    & 0.034407                                 & 0.038086                                         \\ \hline
\end{tabular}
\end{table}

Table 7 represents the sensitivity of $Re_{ss}$ and $\theta_{Avg,ss}$ to $Co_B$. The values of $Co_B$, $Re_{ss}$ and $\theta_{Avg,ss}$ are obtained from Figure 17b. It is observed from Table 7 that a $1\%$ change in $Co_B$ results in a maximum of $\approx 0.01\%$ change in the magnitudes of $Re_{ss}$ and $\theta_{Avg,ss}$, indicating a negligible dependence of steady state parameters on $Co_B$.

\begin{table}[!htb]
\centering
\caption{Sensitivity of $Re_{ss}$ and $\theta_{Avg,ss}$ to $Co_B$}
\begin{tabular}{|l|l|l|l|l|l|l|l|}
\hline
$Co_B$ & $Re_{ss}$ & $\theta_{Avg,ss}$ & $\% \Delta Co_B$ & $\% \Delta Re_{ss}$ & $\% \Delta \theta_{Avg,ss}$ & $\frac{\% \Delta Re_{ss}}{\% \Delta Co_B}$ & $\frac{\% \Delta \theta_{Avg,ss}}{\% \Delta Co_B}$ \\ \hline
0.0005 & 15.50876  & 0.016207          &                &                     &                             &                                          &                                                  \\ \hline
0.005  & 15.42978  & 0.016102          & 900            & -0.50924            & -0.64521                    & -0.00057                                 & -0.00072                                         \\ \hline
0.05   & 14.87981  & 0.015375          & 900            & -3.56438            & -4.51566                    & -0.00396                                 & -0.00502                                         \\ \hline
0.5    & 13.75766  & 0.013891          & 900            & -7.54143            & -9.65029                    & -0.00838                                 & -0.01072                                         \\ \hline
\end{tabular}
\end{table}

Table 8 represents the sensitivity of $Re_{ss}$ and $\theta_{Avg,ss}$ to $St$. The values of $St$, $Re_{ss}$ and $\theta_{Avg,ss}$ are obtained from figure 18b corresponding to $Co_B=0.5$. It is observed from Table 8 that a $1\%$ change in $St$ results in a maximum of $\approx 0.01\%$ change in the magnitude of $Re_{ss}$ and $\approx 0.1\%$ change in the magnitude of $\theta_{Avg,ss}$, indicating the negligible dependence of $Re_{ss}$ and a dependence of $\theta_{Avg,ss}$ on $St$.

\begin{table}[!htb]
\centering
\caption{Sensitivity of $Re_{ss}$ and $\theta_{Avg,ss}$ to $St$}
\begin{tabular}{|l|l|l|l|l|l|l|l|}
\hline
$St$  & $Re_{ss}$ & $\theta_{Avg,ss}$ & $\% \Delta St$ & $\% \Delta Re_{ss}$ & $\% \Delta \theta_{Avg,ss}$ & $\frac{\% \Delta Re_{ss}}{\% \Delta St}$ & $\frac{\% \Delta \theta_{Avg,ss}}{\% \Delta St}$ \\ \hline
100   & 14.0544   & 0.023575          &                &                     &                             &                                          &                                                  \\ \hline
400   & 13.95849  & 0.015829          & 300            & -0.68243            & -32.8548                    & -0.00227                                 & -0.10952                                         \\ \hline
700   & 13.83432  & 0.014498          & 75             & -0.88952            & -8.40748                    & -0.01186                                 & -0.1121                                          \\ \hline
1,000 & 13.75766  & 0.013891          & 42.85714       & -0.55416            & -4.18764                    & -0.01293                                 & -0.09771                                         \\ \hline
\end{tabular}
\end{table}

\section*{\hfil \Large Nomenclature \hfil}

\begin{tabular}{ll}
	
	$A_{cs}$ &  Area of the cross-section ($m^2$)\\
	$C_{p}$ & Specific heat capacity ($J/KgK$)\\
	$D_h$ & Hydraulic diameter of both loop 1 \&\ 2 ($m$)\\
	$g$ & Gravitational constant ($m/s^2$)\\
	$L$ & CNCL height used for 1-D model ($m$)\\
	$L1$ &  CNCL width used for 1-D model ($m$)\\
	$m_1$ &  Magnitude of linear slope of $\theta$ in the common heat exchange section of Loop-1\\
	$m_2$ &  Magnitude of linear slope of $\theta$ in the wall\\
	$m_3$ &  Magnitude of linear slope of $\theta$ in the common heat exchange section of Loop-2\\
	$N$ & Number of bends on the component NCL of the CNCL system\\
	$Q^{\prime\prime}$ & Heat flux ($W/m^2$)\\
	$R$ &  Radius of curvature of the bend ($m$)\\
	$T_{1}$ & Temperature of Loop 1 ($K$)\\
	$T_{2}$ & Temperature of Loop 2 ($K$)\\
	$t$ &  Time ($s$)\\
	$t_w$ &  Wall thickness ($m$)\\
	$T_{0}$ & Reference temperature of loop 1 \&\ 2 ($K$)\\
	$U$ & Heat transfer coefficient at the heat exchanger wall ($W/m^2K$)\\
	$x$ &  Distance from origin '$O$'($m$)\\
	
\end{tabular}

\section*{Greek letters}

\begin{tabular}{ll}
    $\alpha$ & Thermal diffusivity ($m^2/s$)\\
	$\beta$ & Coefficient of thermal expansion ($1/K$)\\
	$\kappa$ & Thermal conductivity ($W/(mK)$)\\
	$\nu$ & Kinematic viscosity ($m^2/s$)\\
	$\rho$ & Density ($kg/m^3$)\\
	$\rho_0$ & Reference density ($kg/m^3$)\\
	$\rho_w$ & Density of Wall ($kg/m^3$)\\
	$\tau$ & Wall shear stress exerted on fluid ($Pa$)\\
	$\omega_{1}$ &  Fluid velocity of loop 1 ($m/s$)\\
	$\omega_{2}$ &  Fluid velocity of loop 2 ($m/s$)\\
\end{tabular}

\section*{Non-dimensional numbers}

\begin{tabular}{ll}
	$f_{F}$ & Fanning friction factor ($f_{F}=\tau_{i}/(\frac{1}{2}\rho\omega_{i}^2)=b/Re^d$)\\
	$Gr_m$ & Modified Grashof number ,Vijayan (2002) \cite{vijayan2002} \\
	$K$ & 	Bend losses coefficient  ($K=\Delta P_{bend}/(\frac{1}{2}\rho\omega_{i}^2)$)\\
	$N_g$ & Geometric parameter ,Vijayan (2002) \cite{vijayan2002} \\
	$Pr$ & Prandtl number , ($Pr=\nu/a$)
\end{tabular}

\section*{Constants}

\begin{tabular}{ll}
	$b$ & 14.23 (for fully developed flow in laminar regime)\\
	$d$ & 1 (for fully developed flow in laminar regime) \\
	$\Delta T$ & ( $\Delta T_i=(4Q^{\prime\prime}t_0)/(\rho_i Cp_i D_h) $) \\
	$\Delta T_w$ & ( $\Delta T_i=(4Q^{\prime\prime}t_0)/(\rho_w Cp_w t_w) $) \\
	$to$ & ( $t_0={x_0D_h}/{\nu_1} $) \\
	$x_0$ & ( $x_0=(L+L1)$) \\
\end{tabular}

\section*{Piece-wise functions}
\begin{tabular}{ll}
	
	$f(x)$ & Function which represents the geometry of the loop\\
	$h_1 (x)$ & Function which represents the heating section location\\
	$h_2 (x)$ & Function which represents the cooling section location \\
	$\lambda (x)$ & Function which represents the location of thermal coupling on the CNCL \\
	$j_1 (x)$ &  $h_1 (x)=\frac{4Q^{\prime \prime}}{\rho C_p D_h}  j_1 (x) $ \\
	$j_2 (x)$ &  $h_2 (x)=\frac{4Q^{\prime \prime}}{\rho C_p D_h}  j_2 (x) $ \\
\end{tabular}

\section*{Fourier coefficients}
\begin{tabular}{ll}
	
	$\alpha_k$ & Fourier coefficient of $T_1(x,t)$\\
	$\beta_k$ & Fourier coefficient of $T_2(x,t)$ \\
	$\gamma_k$ & Fourier coefficient of $h_1(x)$ \\
	$\delta_k$ & Fourier coefficient of $h_2(x)$ \\
	$\zeta_k$ & Fourier coefficient of $\lambda(x)$ \\
	$A_k$ & Fourier coefficient of $f(x)$ \\
\end{tabular}

\section*{Non dimensional parameters}

\begin{tabular}{ll}
    $Co_A$ & $Co_A$=$Co_1$ when same fluids are used in Loop-1 and Loop-2\\
    $Co_B$ & $Co_B$=$Co_2$=$Co_4$ when same fluids are used in Loop-1 and Loop-2\\
	$Co_1$ & Flow resistance coefficient  ($Co_1=\frac{2bx_0}{D_h}$)\\
	$Co_2$ & Thermal coupling sensitivity coefficient between Loop 1 and wall ($Co_2=\frac{\Delta T_w}{\Delta T_1}$)\\
	$Co_3$ & Thermal coupling sensitivity coefficient between Loop 1 and Loop 2  ($Co_3=\frac{\Delta T_2}{\Delta T_1}$)\\
	$Co_4$ & Thermal coupling sensitivity coefficient between Loop 2 and wall ($Co_4=\frac{\Delta T_w}{\Delta T_2}$)\\
	$Fo$ & Fourier number ($Fo_i=\frac{a_it_{0,i}}{x_0^2}$)\\
	$Fo_w$ & Wall Fourier number ($Fo_w=\frac{a_w t_{0,i}}{x_0^2}$)\\
	$Gr$ & Grashof number ($Gr_i=\frac{g \beta_i \Delta T_i x_0 D_h t_{0,i}}{(L+L1) \nu_i}$)\\
	$Nu$ & Nusselt number ($Nu=\frac{h D_h}{\kappa}$)\\
	$Re$ & Reynolds number ($Re_i=\frac{\omega_iD_h}{\nu_i}$)\\
	$St$ & Stanton number ($St_i=\frac{Ut_{0,i}}{\rho_i Cp_i D_h}$)\\
	$s$ & Non dimensional length ( $s={x}/{x_0}$)\\
	$\theta$ & Non dimensional temperature ($\theta_i={(T_i-T_0)}/{\Delta T_i}$)\\
	$\theta_p$ & Non dimensional temperature at $s=0$ of Loop-1\\
	$\theta_q$ & Non dimensional temperature at $s=0$ of the wall\\
	$\theta_r$ & Non dimensional temperature at $s=0$ of Loop-2\\
	$\zeta$ & Non dimensional time ($\zeta={t}/{t_0}$)\\
	
\end{tabular}

\section*{Subscripts}

\begin{tabular}{ll}
	$0$ & Any parameter except $t$ and $x$ at time $t=0$ $s$ \\
	$1$ & Any parameter referring to Loop 1 \\
	$2$ &  Any parameter referring to Loop 2 \\
	$ss$ &  Any parameter considered at steady state\\
	$Avg$ & Average value of the parameter\\
	$i$ & Refers to subscript `1' or subscript `2' according to relevance\\
	$w$ & Any parameter referring to the wall\\
\end{tabular}

\section*{Abbreviations}

\begin{tabular}{ll}
	$CFD$ &  Computational Fluid Mechanics \\
	$CNCL$ & Coupled Natural Circulation Loop \\
	$NCL$ & Natural Circulation Loop \\
\end{tabular}

\bibliography{IJHMT.bib}

\end{document}